\documentclass[fleqn,usenatbib]{mnras}

\usepackage{dcolumn}
\usepackage{siunitx}
\usepackage{newtxtext,newtxmath}

\usepackage[T1]{fontenc}

\DeclareRobustCommand{\VAN}[3]{#2}
\let\VANthebibliography\thebibliography
\def\thebibliography{\DeclareRobustCommand{\VAN}[3]{##3}\VANthebibliography}

\def\gax{\mathrel{\raise.3ex\hbox{$>$}\mkern-14mu\lower0.6ex\hbox{$\sim$}}}
\def\lax{\mathrel{\raise.3ex\hbox{$<$}\mkern-14mu\lower0.6ex\hbox{$\sim$}}}
\def\gtorder{\mathrel{\raise.3ex\hbox{$>$}\mkern-14mu
             \lower0.6ex\hbox{$\sim$}}}
\def\ltorder{\mathrel{\raise.3ex\hbox{$<$}\mkern-14mu
             \lower0.6ex\hbox{$\sim$}}}

\usepackage{graphicx}	
\usepackage{amsmath}	
\usepackage{changepage}

\title[A Search for Long Term Variability in ASAS-SN]{Life in the Slow Lane: A Search for Long Term Variability in ASAS-SN}

\author[Petz et al.]{
Sydney Petz,$^{1}$\thanks{E-mail: petz.16@osu.edu}
C. S. Kochanek,$^{1,2}$
\\
$^{1}$Department of Astronomy, The Ohio State University, 140 W 18th Avenue, Columbus, OH 43210, USA\\
$^{2}$Center for Cosmology and AstroParticle Physics, 191 W Woodruff Avenue, Columbus, OH 43210, USA\\
}

\date{Accepted XXX. Received YYY; in original form ZZZ}

\pubyear{2015}

\begin{document}
\maketitle

\begin{abstract}
We search a sample of 9,361,613 isolated sources with 13<g<14.5 mag for slowly varying sources.
We select sources with brightness changes larger than $\sim 0.03$~mag/year over 10 years, removing false positives due to, for example, nearby bright stars or high proper motions.
After a thorough visual inspection, we find 782 slowly varying systems. 
Of these systems, 433 are identified as variables for the first time and 349 are previously classified as variables. 
Previously classified systems were mostly identified as semi-regular variables (SR), slow irregular variables (L), spotted stars (ROT), or unknown (MISC or VAR), as long time scale variability does not fit into a standard class.
The stellar sources are scattered across the CMD and can be placed into 5 groups that exhibit distinct behaviors. 
The largest groups are very red subgiants and lower main sequence stars.
There are also a small number of AGN.
There are 551 candidates ($\sim$~70 percent) that also show shorter time scale periodic variability, mostly with periods longer than 10 days.
The variability of 191 of these candidates may be related to dust.

\end{abstract}




\section{Introduction}
There are an increasing number of time domain surveys such as the All-Sky
Automated Survey for SuperNovae (ASAS-SN; \citealt{ Shappee2014, Kochanek2017, Jayasinghe2018}), the Zwicky Transient Facility (ZTF; \citealt{Bellm2014}), the Asteroid
Terrestrial-impact Last Alert System (ATLAS; \citealt{Heinze2018}) and soon the Legacy Survey of Space and Time at the Vera Rubin Observatory (LSST; \citealt{Hambleton2023}).
These surveys are largely focused on the transient universe and variable stars, usually only considering long term variability when looking at active galactic nuclei (AGN).
These surveys (as well as earlier surveys) have never systematically explored the slowly varying universe.
Known classes of longer term variability include magnetic activity cycles on main sequence stars \cite{Baliunas1990}, active K giants \citep{Philips78}, and some classes of young stellar objects (YSOs, \citealt{Teixeira18}).
There are also rarer types including stellar mergers \citep{Tylenda2011} and peculiar stellar dimmings \citep{Simon2018}.

Some of the first explorations of slow variability used sporadic measurements of solar-type stars over several decades \citep{Baliunas1990}. Much like our Sun, they showed cyclic variations in brightness that correspond to magnetic activity cycles.
Magnetic activity and its relation to variability has also been studied in M dwarfs, finding links between brief dips and increases in brightness, rotational spot modulation, and flaring activity \citep{Weis1994}. 
There are cyclic variations on year timescales and a correlation between stronger photometric fluctuations and position above the main sequence \citep{Hosey2015, Clements2017}.

Slow variability has also been observed in K giants on longer timescales. 
Proposed explanations include obscuration from ejected dust shells, lithium flashes, or unstable helium shell burning as these stars begin to ascend up the AGB \citep{Tang2010}. 
The irregular brightening and dimming of K giants may also have a magnetic component \citep{Olah2014}, with changes in the spots and faculae accompanying changes in radius.

Though more traditionally known for their short-term variability, certain sub-classes of classical T Tauri stars (cTTS) exhibit long-term variability.
These Type III cTTS vary by of tenths of a magnitude over a typical observing season due to large-scale variations in their accretion rates or the circumstellar disk region partially occulting the central star \citep{Grankin2007}. These stars can also exhibit smooth changes in mean brightness on timescales of years.

Stellar mergers (also called luminous red novae) are rare, with an estimated rate of approximately one per year in the Galaxy \citep{Kochanek2014}.
After the "classical" transient associated with their peak luminosity, they are observed to slowly evolve for decades.
Interestingly, some systems including V1309 Scorpii and M31-LRN-2015 were observed to show steady variability changes years before their mergers \citep{Tylenda2011, Blagorodnova2020}, and there are theoretical reasons to expect that all of these systems do so \citep{Metzger17, Matsumoto22}. 
This has led to attempts to identify systems which might be in this pre-merger phase (e.g., \citealt{Molnar2017, Addison2022}).

\begin{figure*}
    \includegraphics[width=\columnwidth]{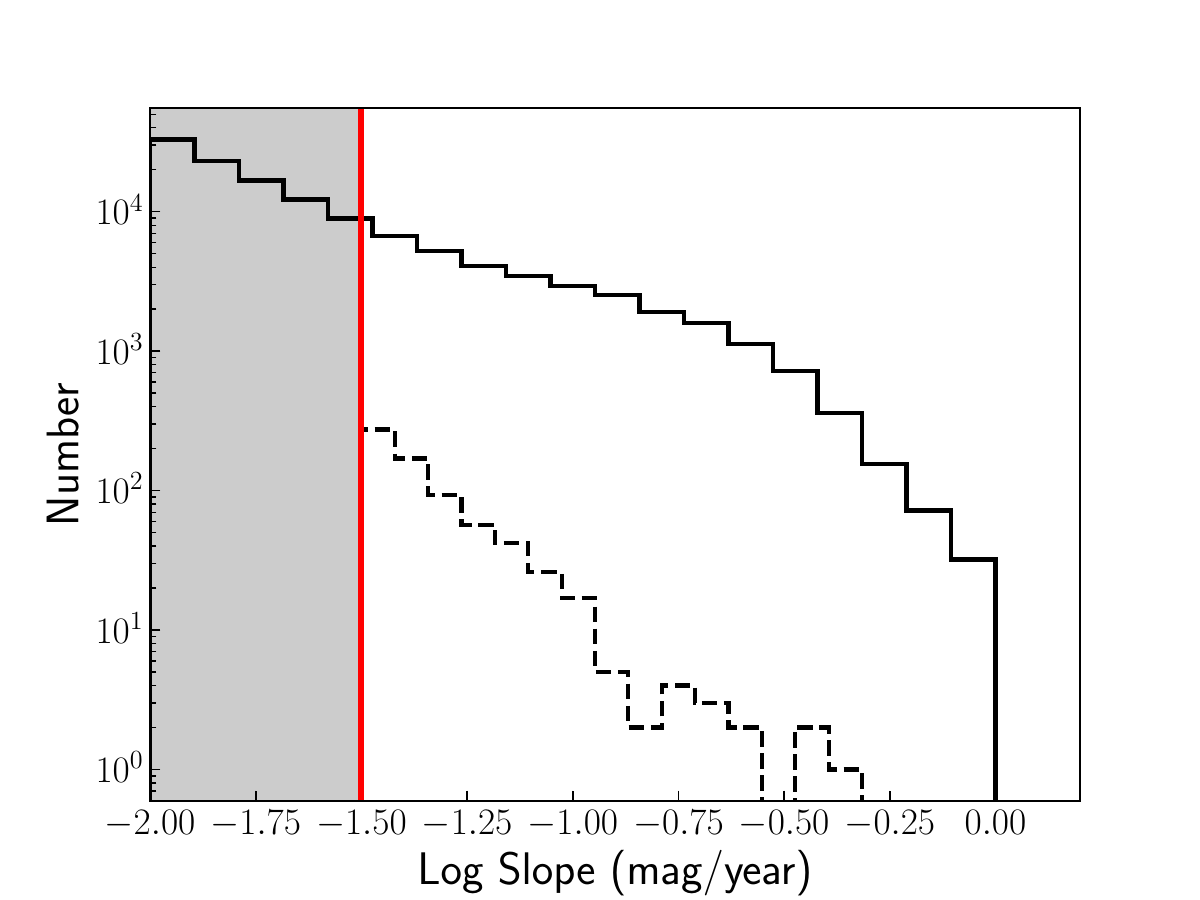}
     \includegraphics[width=\columnwidth]{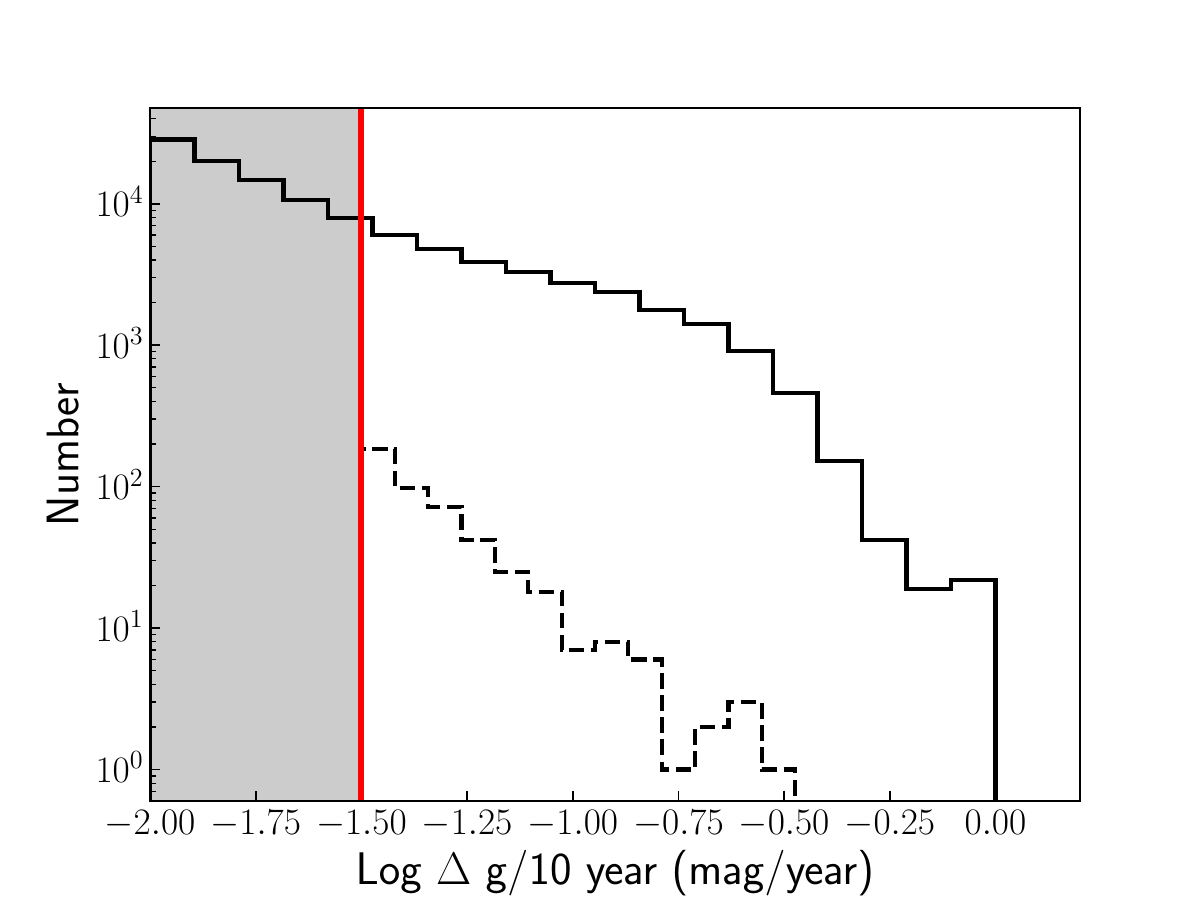}
    \caption{Distributions of the input sources (solid) in linear slope (left) and maximum magnitude change $\Delta g$ in the quadratic fit (right). We considered sources
    to the right of the red lines with slopes or $\Delta g/10$~year $>0.03$~mag/year. The dashed histograms
    show the distributions of the final sample after 
    eliminating false positives.
    }
    \label{fig:1dslopecuts}
\end{figure*}

\begin{figure}
    \includegraphics[width=\columnwidth]{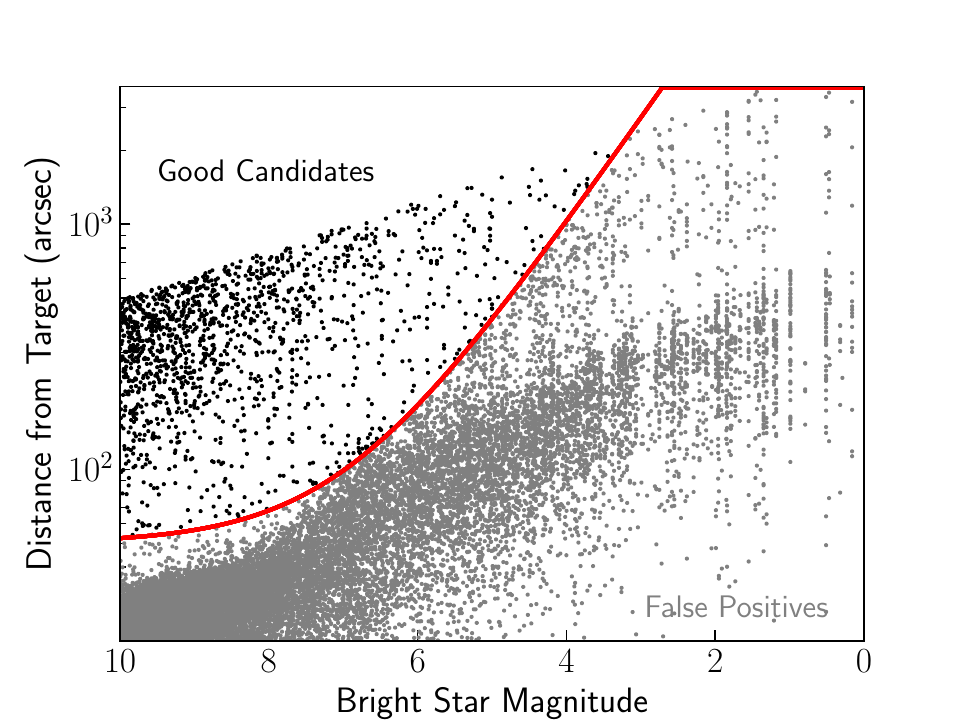}
    \caption{Distribution of the initial candidates in the magnitude and distance of bright nearby stars. We keep stars above the red curve.}
    \label{fig:bscuts}
\end{figure}

There are several examples of slow variability that have yet to be fully explained. 
One example is the main sequence F star KIC 8462852, which has exhibited an unusual series of brief dimming events resulting in a cumulative fading of 3 percent \citep{Montet2016}. Although the individual dimming events could be attributed to polar spots over a period of years or transiting circumstellar material, the long-term variability is not well explained.
A more dramatic example is the cool giant Gaia17bpp which exhibited $\sim 4.5$ magnitude dimming event over a period of more than 6.5 years \citep{Tzanidakis2023}. Though not the only dimming event observed over such a long interval (e.g., \citealt{Rowan2021,Smith2021, Torres2022}), it is one of the deepest and longest known events. Although proposed to be a binary star system with a secondary star enshrouded by an optically thick debris disk \citep{Tzanidakis2023}, the nature of this system is not fully understood, as the secondary star has yet to be identified.
A main-sequence F star, known as Boyajian's star \citep{Boyajian2016}, has exhibited unusual brightening episodes followed by a steady dimming since 2013 \citep{Simon2018}. Multiple theories for these events have been suggested, including the ingestion of a planet and its satellite system \citep{Metzger2017} and energy from the convective structure of the star \citep{Foukal2017}.
Another interestingly varying source is KH 15D, a pre-main sequence star with a history of variation that dates back ~60 years. This weakly accreting T Tauri binary has exhibited photometric variations likely linked to a precessing circumbinary disk and scattered light \citep{Herbst2010}.
The rise and fall of Sakurai's object (V4334 Sgr) marks another dramatic example of slow variation, with a helium flash causing a several magnitude increase in brightness followed by an approximately 11 magnitude decline over several years due to growing dust obscuration \citep{Duerbeck2000}.

Using 11 years of monitoring data, \cite{Neustadt2021} observed 26 nearby galaxies with the Large Binocular Telescope \citep{Kochanek2008, Gerke2015} to search for failed supernovae.
During this search, they identified a peculiar population of luminous stars slowly changing in brightness by factors of $\sim 3$ at nearly constant color.
These stars were too faint to easily characterize, but we can potentially identify Galactic analogues. 
With approximately 10 years of observations, ASAS-SN has obtained photometric measurements of $\sim$100 million stars with $g \lesssim 18.5$ mag. 
Here we use the ASAS-SN data to search for slowly varying sources.
We describe the methods in \S\ref{sec:methods} and the results in \S\ref{sec:Results}.
We discuss possible follow up studies and survey extensions in \S\ref{sec:conclusions}.



\begin{figure*}
    \includegraphics[width=2\columnwidth]{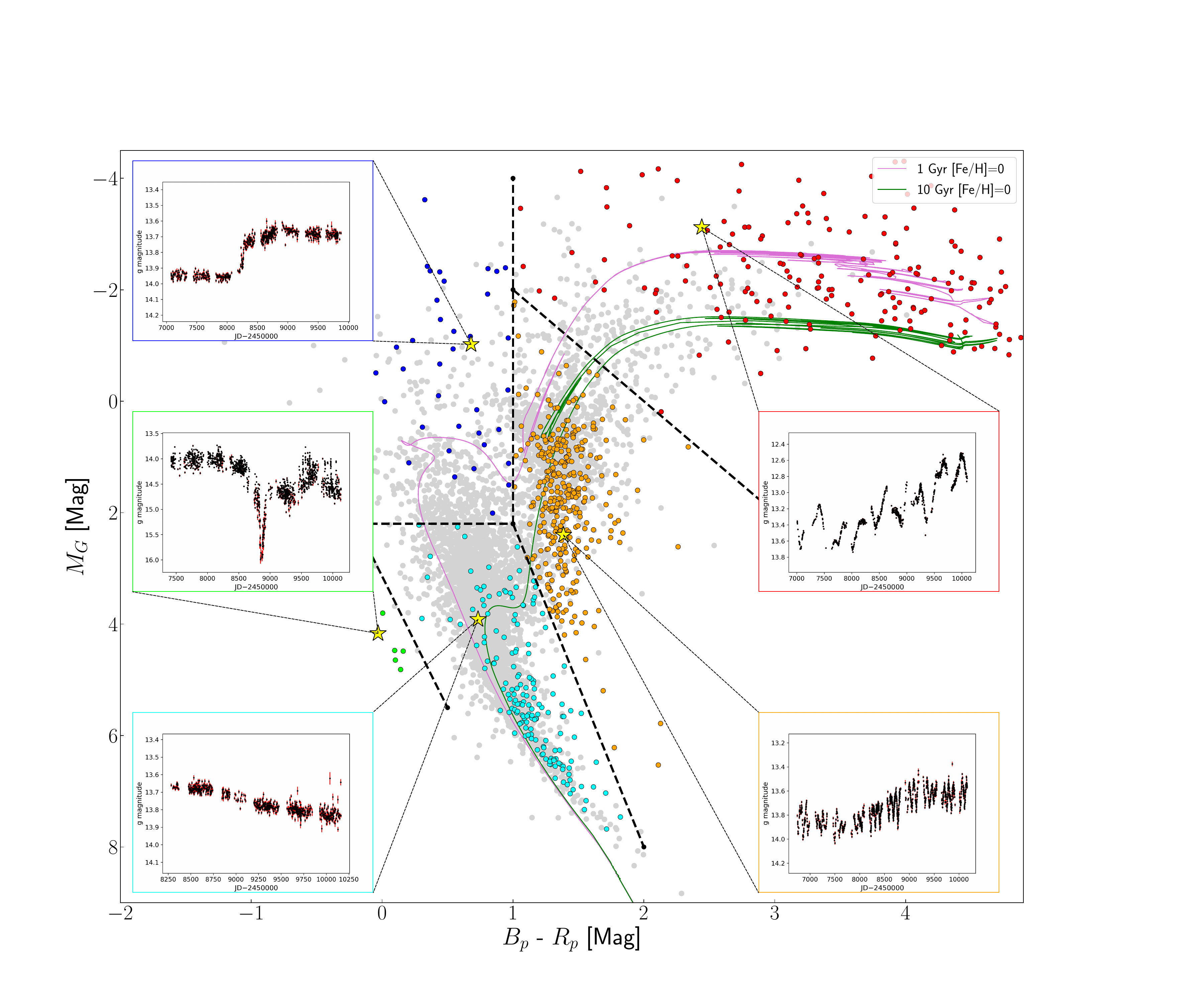}
    \caption{The Gaia DR3 $M_{G}$ and $B_{P} - R_{P}$ color-magnitude diagram of the final candidates divided into groups based on the dashed lines. The curves are solar metallicity 1 and 10 Gyr MIST isochrones.
    One type of light curve is shown in each group.}
    \label{fig:CMDgroups}
\end{figure*}

\begin{figure*}
    \includegraphics[width=2\columnwidth]{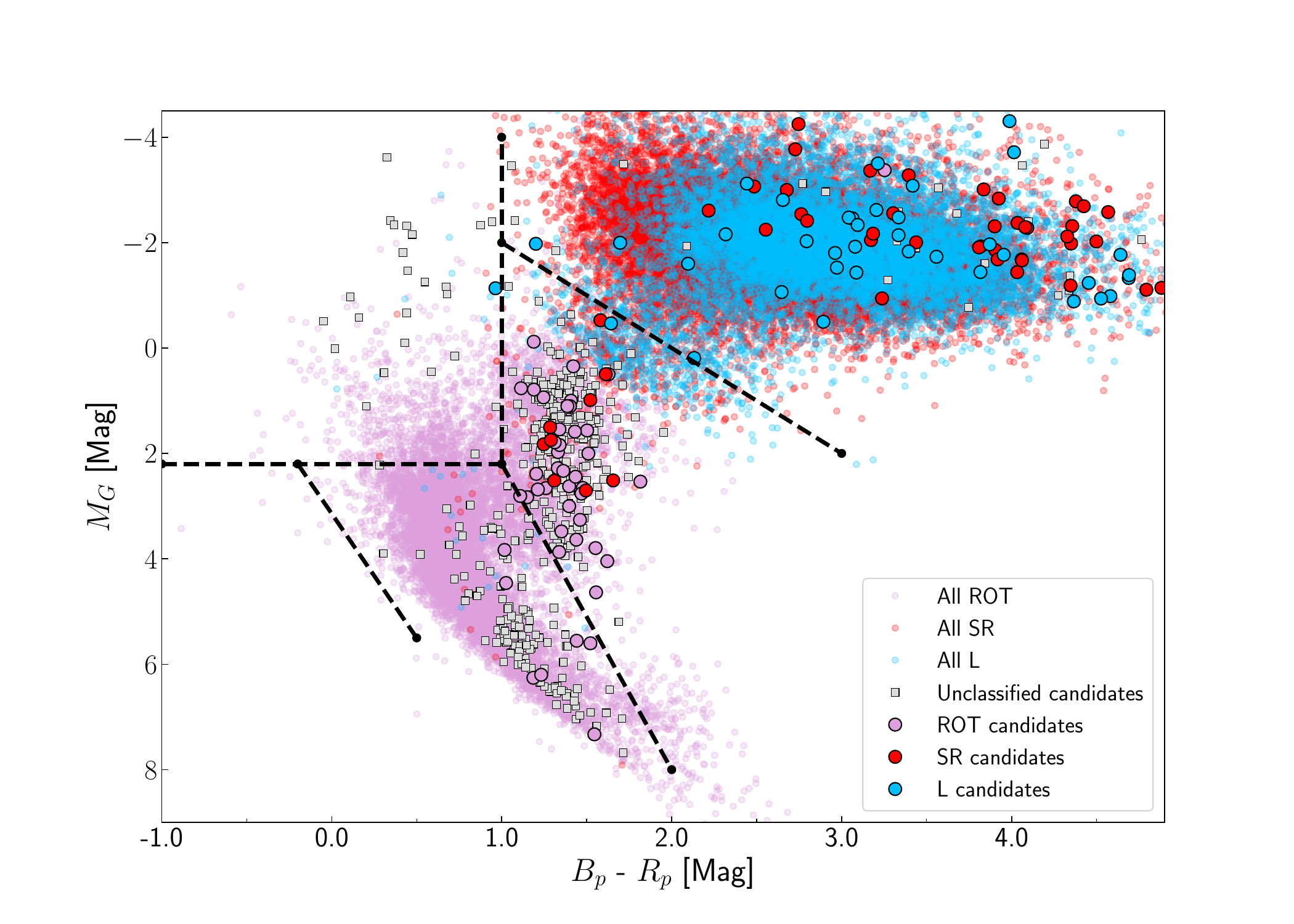}
    \caption{The Gaia DR3 $M_{G}$ and $B_{P} - R_{P}$ color-magnitude diagram of all variables in SkyPatrol V2.0 with 13<g<14.5 that are classified as SR, L, and ROT (small points) as compared to the candidates with these classifications (large points). Previously unclassified candidates are marked by squares.}
    \label{fig:twincmd}

\end{figure*}

\section{Methodology}
\label{sec:methods} 
We first downloaded the light curves of the 9,361,613 sources with 13<g<14.5 mag from ASAS-SN Sky Patrol V2.0 \citep{Hart2023}. 
We then intercalibrated the cameras and both the V and g band data using a damped random walk Gaussian process for interpolation with the camera/filter calibration shifts fit as additional linear parameters 
(see \citealt{Kozlowski2010}). 
Next, we computed seasonal medians to eliminate all short time scale variability. We use the median rather than the mean as it is insensitive to outliers.
We then fit the seasonal medians using both a linear and a quadratic function of time. Fig~\ref{fig:1dslopecuts} shows the distribution of linear slopes. 
We kept candidates with linear slopes greater than 0.03 mag/year.
We used the quadratic fits to estimate the maximum implied difference in magnitude $\Delta g$ over the length of each light curve. We kept stars with a $\Delta g$ greater than 0.3 magnitudes.
With roughly $\Delta t$ $\approx$ 10 years of data, a $\Delta g$ > 0.3 mag implies a "slope" $\Delta g$ / $\Delta t\gtrsim$ 0.03~mag/year, as also shown in Fig~\ref{fig:1dslopecuts}
This left us with 36,705 candidates.

Among the initial candidates, we find four classes of false positives. 
The first class is a range of artifacts created by bright stars.
Because of the bleed trails of saturated stars, the star causing the false positives can be quite distant. Figure~\ref{fig:bscuts} shows all the initial candidates with absolute linear slopes greater than $0.03$~mag/year in the space of the distance and magnitude of nearby bright stars.
We identify two bands, with the lower band consisting of false positives due to the nearby bright stars. 
To discard these false positives, we keep the stars above the quadratic curve shown in Figure~\ref{fig:bscuts} up to a maximum distance of 3600 arcseconds at brighter magnitudes.
This reduces our sample to 6690 stars.

The second class of false positives is stars near the celestial south pole. Due to field rotations created  on ASAS-SN's equatorial mounts, there is a concentration of stars with nonphysical maximum magnitude differences at this pole. To account for these, we discard  any candidates with declinations less than $-$88 degrees. This cuts 75 additional stars from our sample.

The third class of false positives are high proper motion stars or stars near high proper motion stars.
While ASAS-SN photometric apertures are fairly large (radius of 16"), stars with sufficiently large proper motions will create a signal as they move through the aperture. 
Because the resolution is low, sufficiently bright non-target stars moving through the aperture will also create a signal. 
To remove these objects from our sample, we discard stars with proper motions higher than 100 mas/year or where a nearby star with a flux ratio greater than 0.01 passes through the photometric aperture.
This removes 49 of the remaining candidates.

We examined the remaining light curves and their corresponding ASAS-SN images individually.
In doing so, we can not only begin to examine the long-term trends of these objects but also remove targets that appear too noisy or are still being affected by nearby bright stars. 
During this process, we discovered a fourth source of false positives, stars with insufficient overlap between the V and g band data where the intercalibration process fails.
This sometimes leads to a magnitude jump, mimicking a brightening event. We discard these targets as a part of our visual inspection. 
This left us with the 782 candidates listed in Table~\ref{tab:cands}.
Their distribution in slope and $\Delta g$ is shown in Fig.~\ref{fig:1dslopecuts}.
One useful initial check for systematic problems is that there are roughly equal numbers of sources becoming brighter (395) and fainter (387), so they are not affected by systematic drifts in the photometry.

We cross matched these sources with 
Gaia Alerts \citep{GaiaAlerts}, and the AAVSO VSX catalog \citep{AAVSO}, which contains variables identified by amateurs and surveys such as ASAS-SN \citep{Jayasinghe2020, Jayasinghe2021, Christy2023}, ATLAS \citep{Heinze2018}, WISE \citep{Chen2018}, and ZTF \citep{Chen2020}.
We match our final candidates with Gaia DR3 photometry \citep{GaiaDR3}, use distances from \cite{BJ2023}, and correct for extinction with \texttt{mwdust} \citep{Mwdust}, which is based on the dust maps of \cite{Drimmel}, \cite{Marshall2006} and \cite{Green2019}. We identified likely active galactic nuclei (AGN) by matching the sources to the {\tt milliquas v8} catalog (\citealt{Flesch2023}).  We used {\tt SIMBAD} (\citealt{Wenger2000}) to obtain general stellar classifications and spectral types if they were available.
Because many of the candidates appear to be periodic on shorter time scales, we fit and subtract a linear or quadratic trend from each light curve to remove the observed long time scale trend and then use a Lomb-Scargle periodogram \citep{Lomb1976,Scargle1982} to search for periodicity.
We discard periods longer than the average observing season and kept the most significant period with a false alarm probability $<$ 0.1
for each candidate to remove nonphysical periods.
We also extract Near-Earth Object Wide-field Infrared Survey Explorer (NEOWISE; \citealt{Mainzer2014}) infrared light curves for each source using \cite{WISE-pull-lc}.
We combine closely spaced points, and fit the W1 light curve and the W1$-$W2 color evolution with linear and a quadratic functions of time.


\section{Results}
\label{sec:Results}
After searching through a sample of 9,361,613 objects within 13<g<14.5 magnitudes, we find a total of 782 candidates exhibiting long term variability.
This includes 349 objects previously flagged as variables, and 433 new variables.
We show all of the candidates in Fig.~\ref{fig:CMDgroups} on a Gaia DR3 $M_{G}$ and $B_{P}-R_{P}$ color magnitude diagram as well as Solar metallicity 1 Gyr and 10 Gyr MIST \citep{Paxton2018} isochrones to track stellar evolutionary stages.
We loosely separate the stellar candidates into five main groups based on the isochrones in Fig.~\ref{fig:CMDgroups}: main sequence, subgiant/giants, AGB stars, luminous blue stars, and novae. We find that the variables in each group usually exhibit different physical characteristics in their light curves.
Due to negative or missing parallaxes and distances in Gaia DR3 \citep{GaiaDR3} and \cite{BJ2023}, 27 candidates cannot be placed into a group, and 4 of these are AGN.

\begin{figure*}
    \includegraphics[width=0.66\columnwidth]{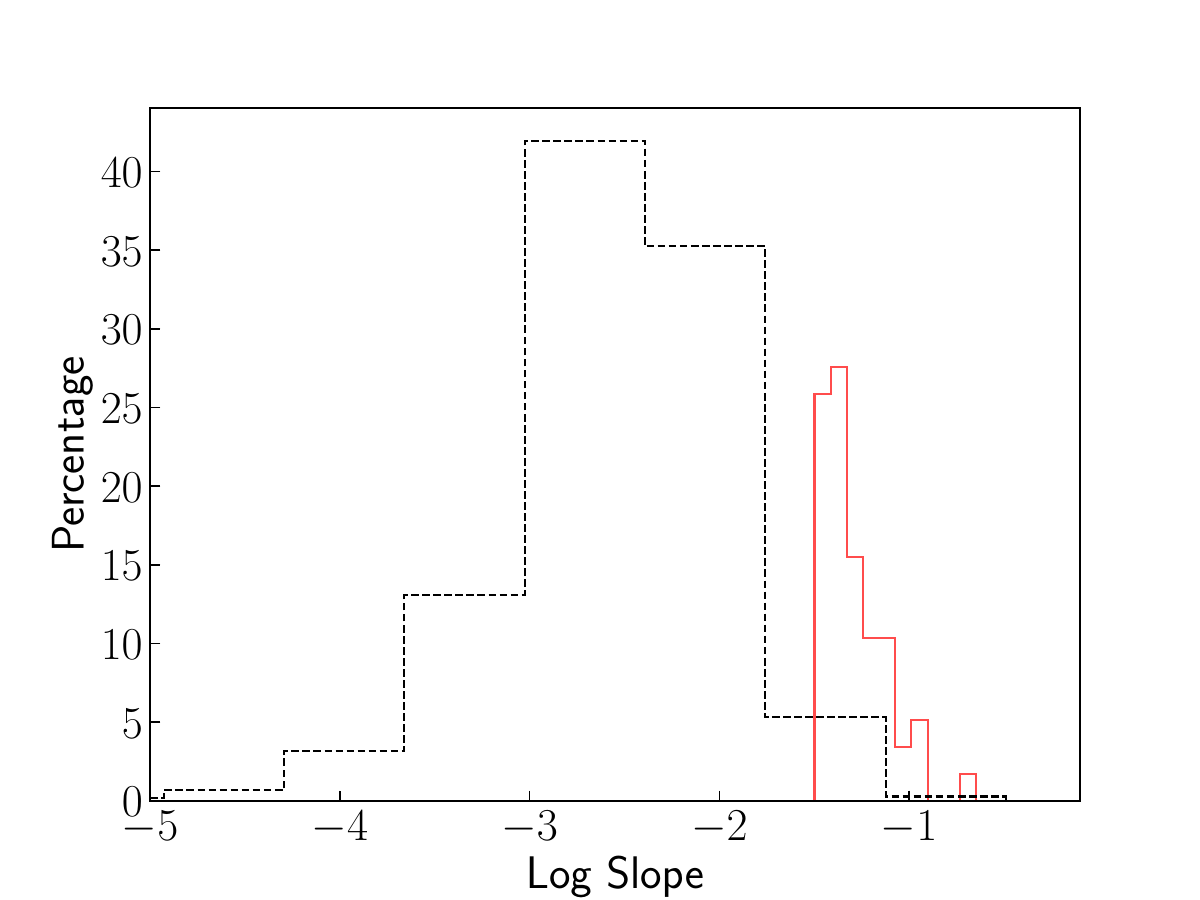}\includegraphics[width=0.66\columnwidth]{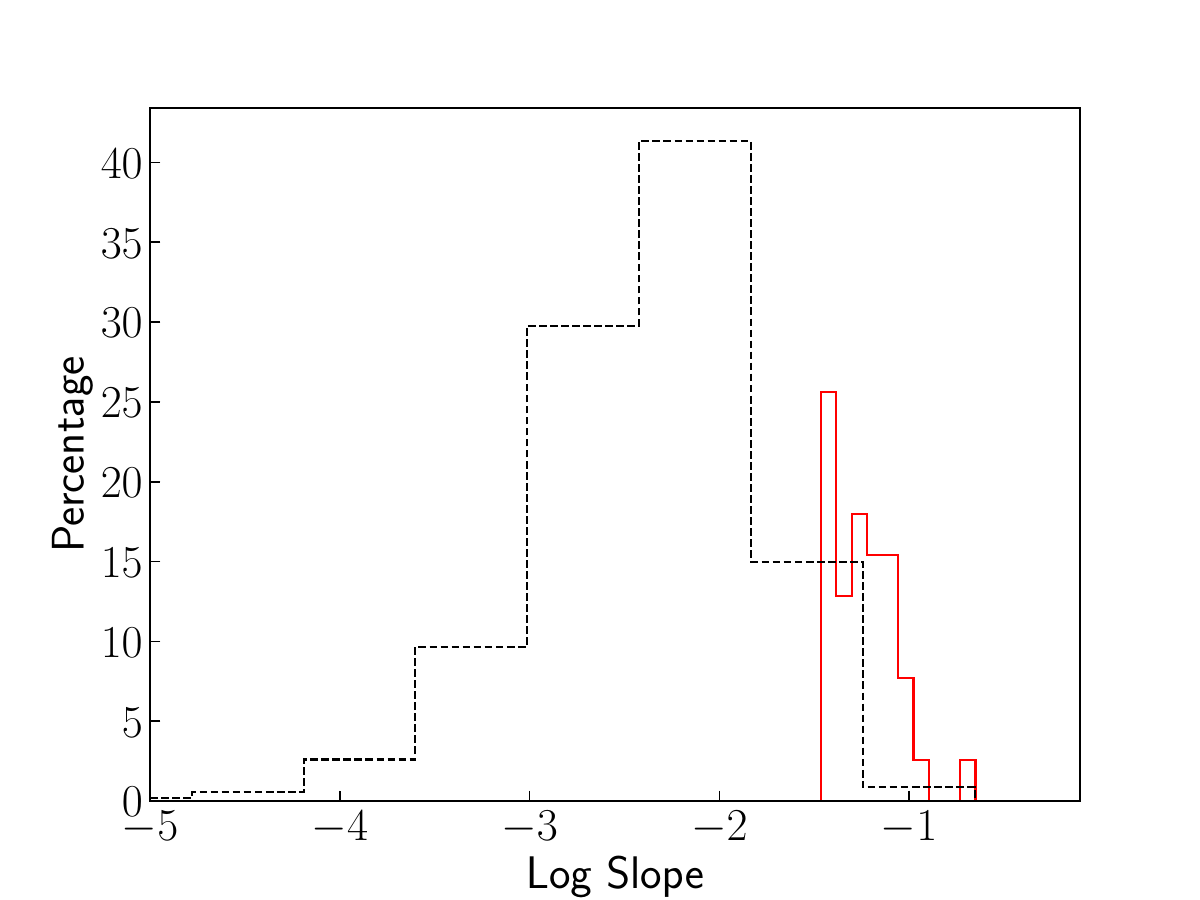}\includegraphics[width=0.66\columnwidth]{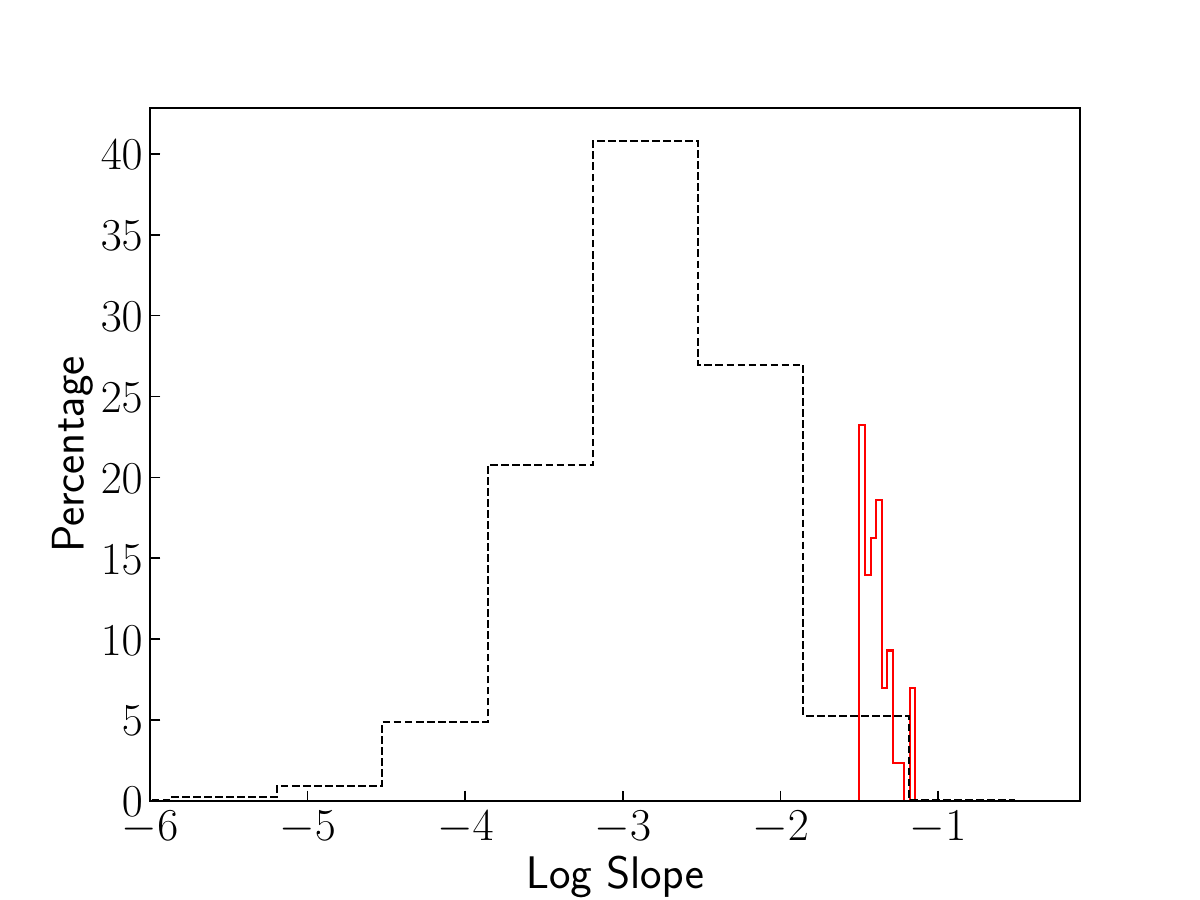}
    \caption{Slope distribution of the candidates (red solid) and all variables (black dashed) in SkyPatrol V2.0 with 13<g<14.5 that are classified as SR (left), L (middle), and ROT (right).}
    \label{fig:twinslopes}

\end{figure*}

\begin{figure}
    \includegraphics[width=1.0\columnwidth]{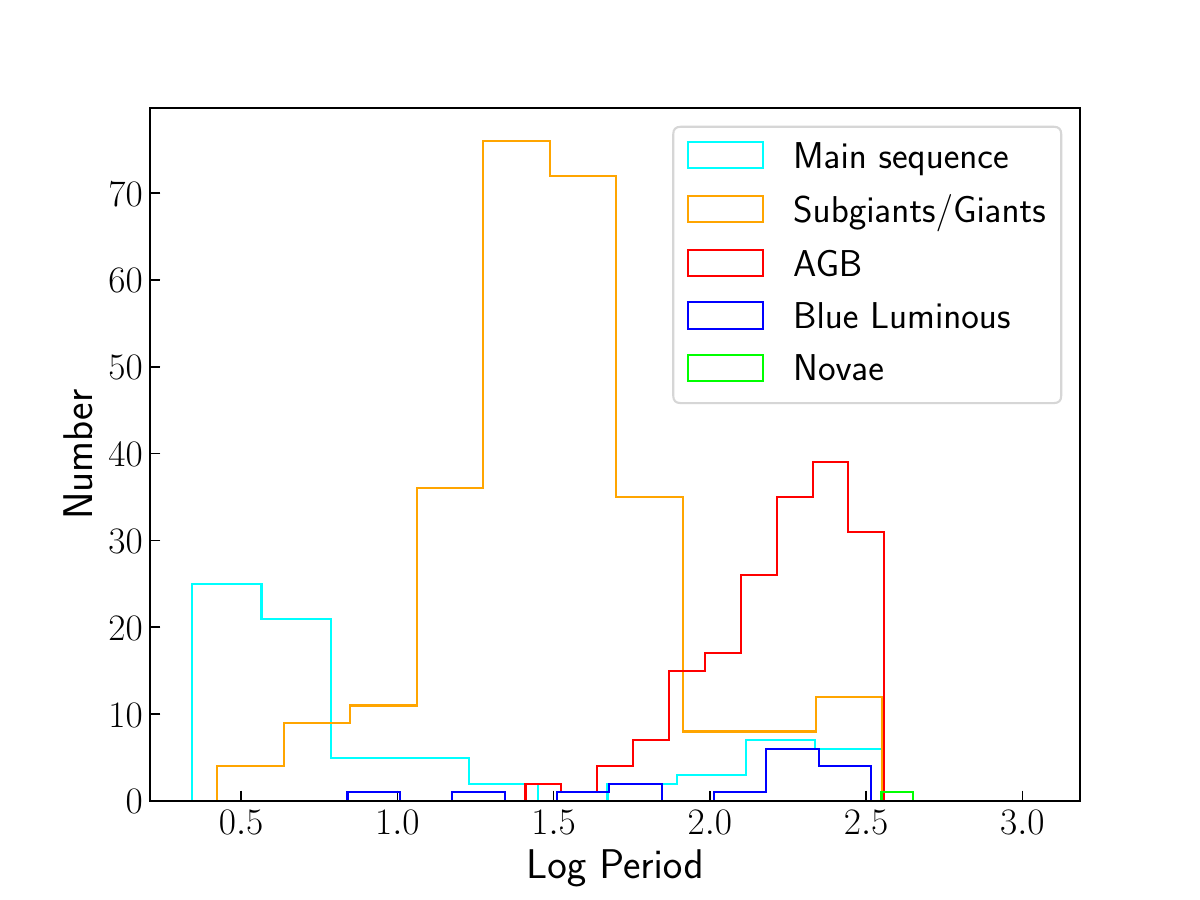}
    \caption{Period distribution for the candidates by group.}
    \label{fig:periodhist}
\end{figure}

\begin{figure*}
    \includegraphics[width=1.0\columnwidth]{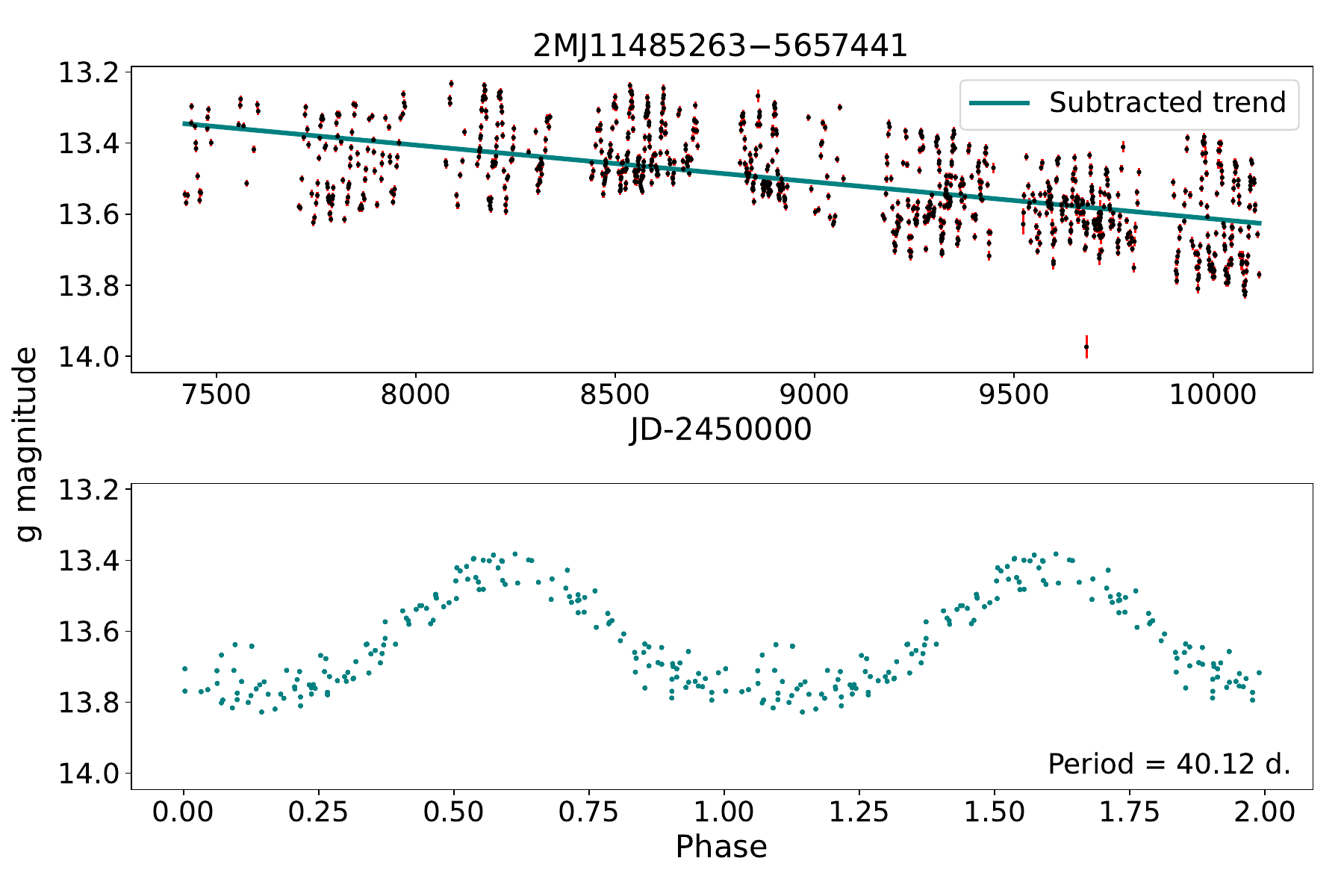}\includegraphics[width=1.0\columnwidth]{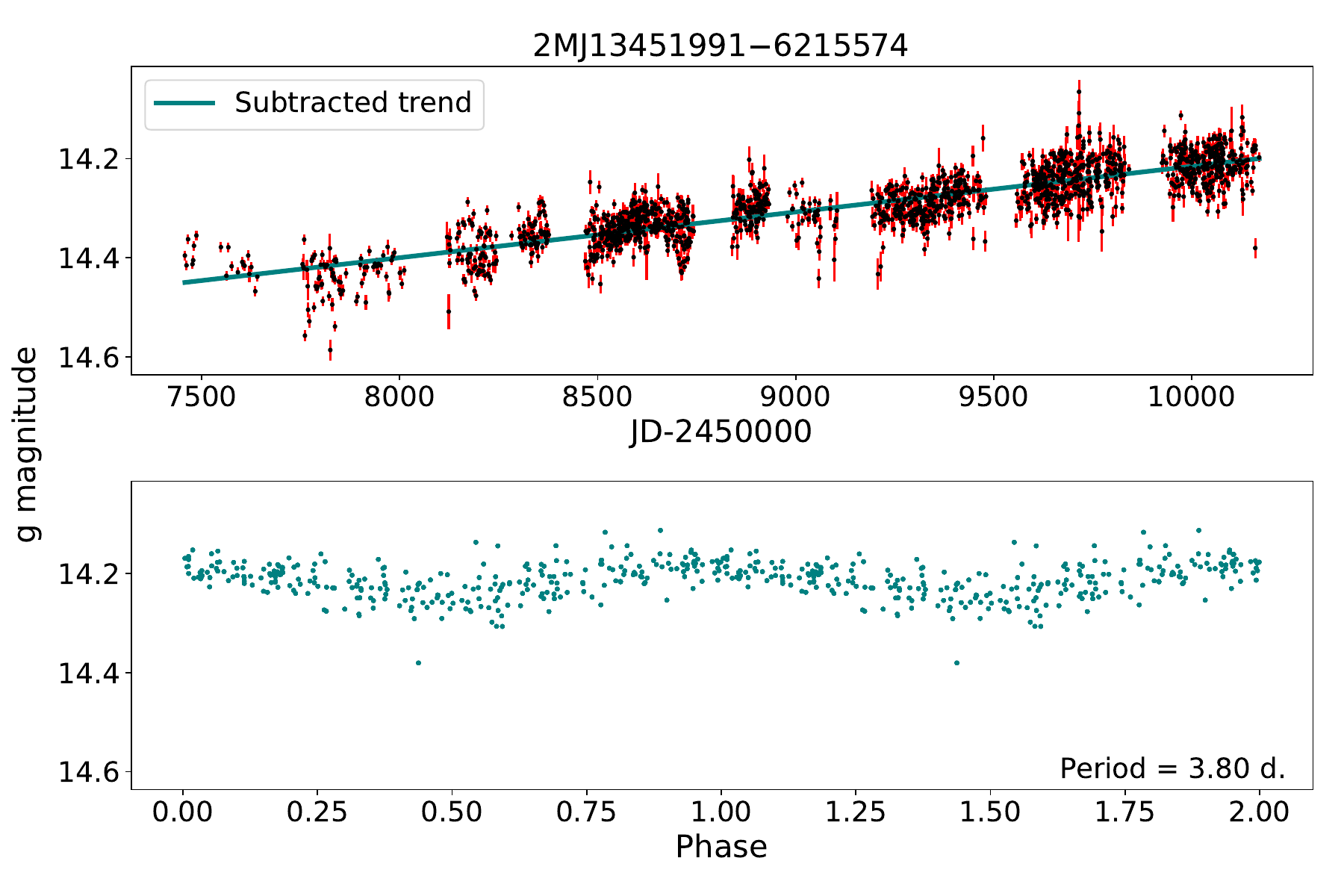}
    \caption{Phased light curves of two candidates displaying regular periodicity, labeled by their 2MASS ID. The top panel shows the light curve and the trend polynomial subtracted before estimating the period. The bottom panel shows the phased, detrended light curve. Left: A subgiant/giant group candidate with a period of 40.12 days. Right: A main sequence group candidate with a period of 3.80 days.}
    \label{fig:periodplots}
\end{figure*}

\begin{figure*}
    \includegraphics[width=0.66\columnwidth]{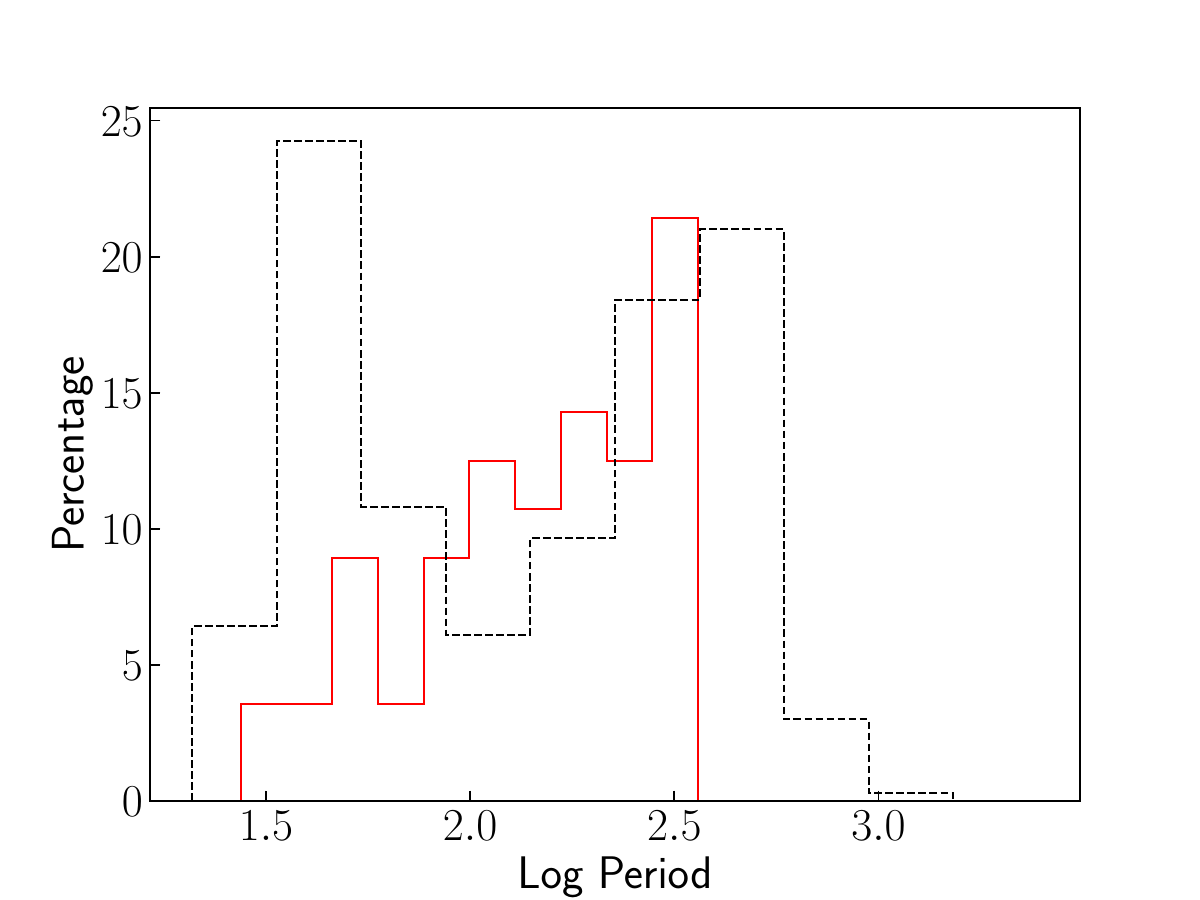}\includegraphics[width=0.66\columnwidth]{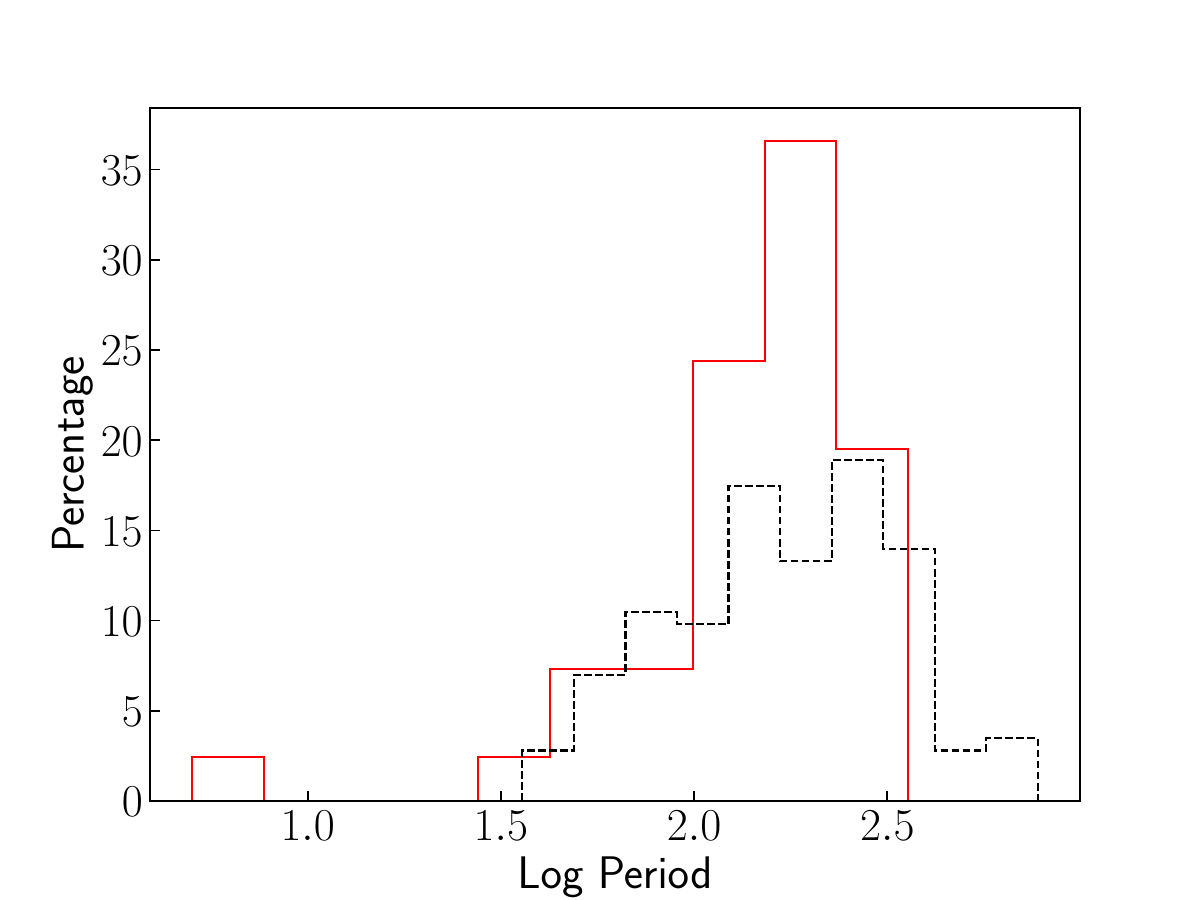}\includegraphics[width=0.66\columnwidth]{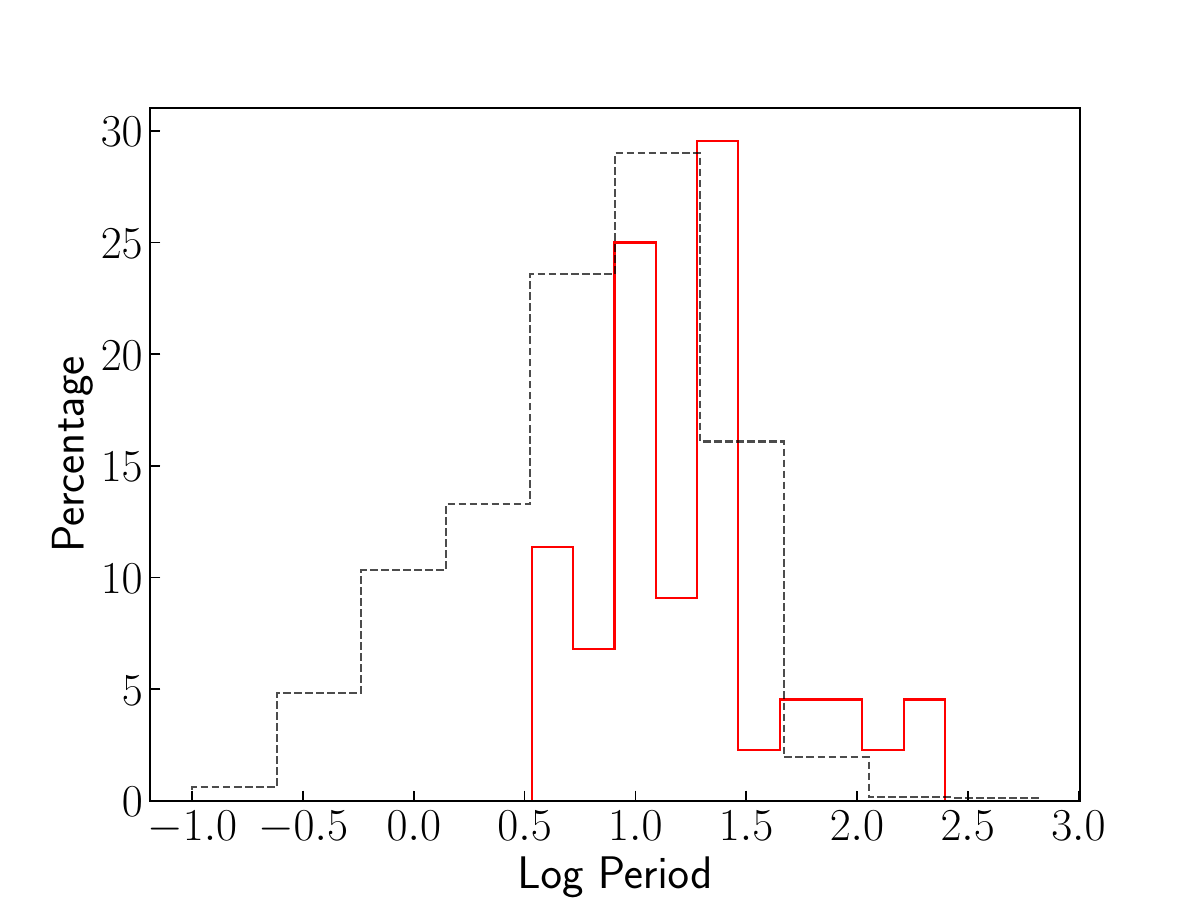}
    \caption{Period distribution of the candidates (red solid) and all variables (black dashed) in SkyPatrol V2.0 with 13<g<14.5 that are classified as SR (left), L (middle), and ROT (right).}
    \label{fig:twinsperiods}
\end{figure*}

The two most common types of known variables in the sample are semi-regular (SR) and slow irregular (L)
pulsating variables, with a total of 112 pulsating variables in all.  There are also 88
classified as a variable of unknown type (MISC or VAR), which is not 
surprising since the slow variability we are searching for is not a standard type.
There are also 13 eclipsing variables, 53 rotating variables, 41 eruptive variables
and 13 cataclysmic variables. Three were assigned to multiple types and 8 were identified as AGN in {\tt milliquas v8}. Fig.~\ref{fig:twincmd} compares the candidates previously classified as SR, L, and ROT candidates to all
such variables in Sky Patrol v2.0 with $13 < g < 14.5$~mag in a Gaia DR3 $M_{G}$ and $B_{P}-R_{P}$ color magnitude diagram and Fig.~\ref{fig:twinslopes}
compares their slope distributions. 
The candidates previously classified as ROT variables are RS CVn or subgiant rotational variables while many of the new systems are on the main sequence. Many of the candidates previously classified as SR variables also appear to be these rapidly rotating (sub)sub-giant stars. The remaining candidates previously classified as SR variables and those classified as L variables appear somewhat more luminous than the typical systems.


Fig.~\ref{fig:periodhist} shows the distribution of the periodic variables in each group, and Fig.~\ref{fig:periodplots} shows two examples with a roughly 
sinusoidal periodic signal on time scales short compared to the long term 
trends (40 days for the one from the giant/subgiant group, 
and 3.8 days for the main sequence example).
Not surprisingly, the periods are basically just ordered by 
stellar size, with the main sequence group at short periods, the subgiant/giants
at intermediate periods, and the AGB group at long periods.  There are few
periodic stars in the luminous blue and nova groups.  Fig.~\ref{fig:twinsperiods}
compares the rotation periods of our candidates previously classified as
SR, L or ROT variables to all such variables with $13 < g < 14.5$
in SkyPatrol V2.0.    

\begin{figure*}
    \includegraphics[width=\columnwidth]{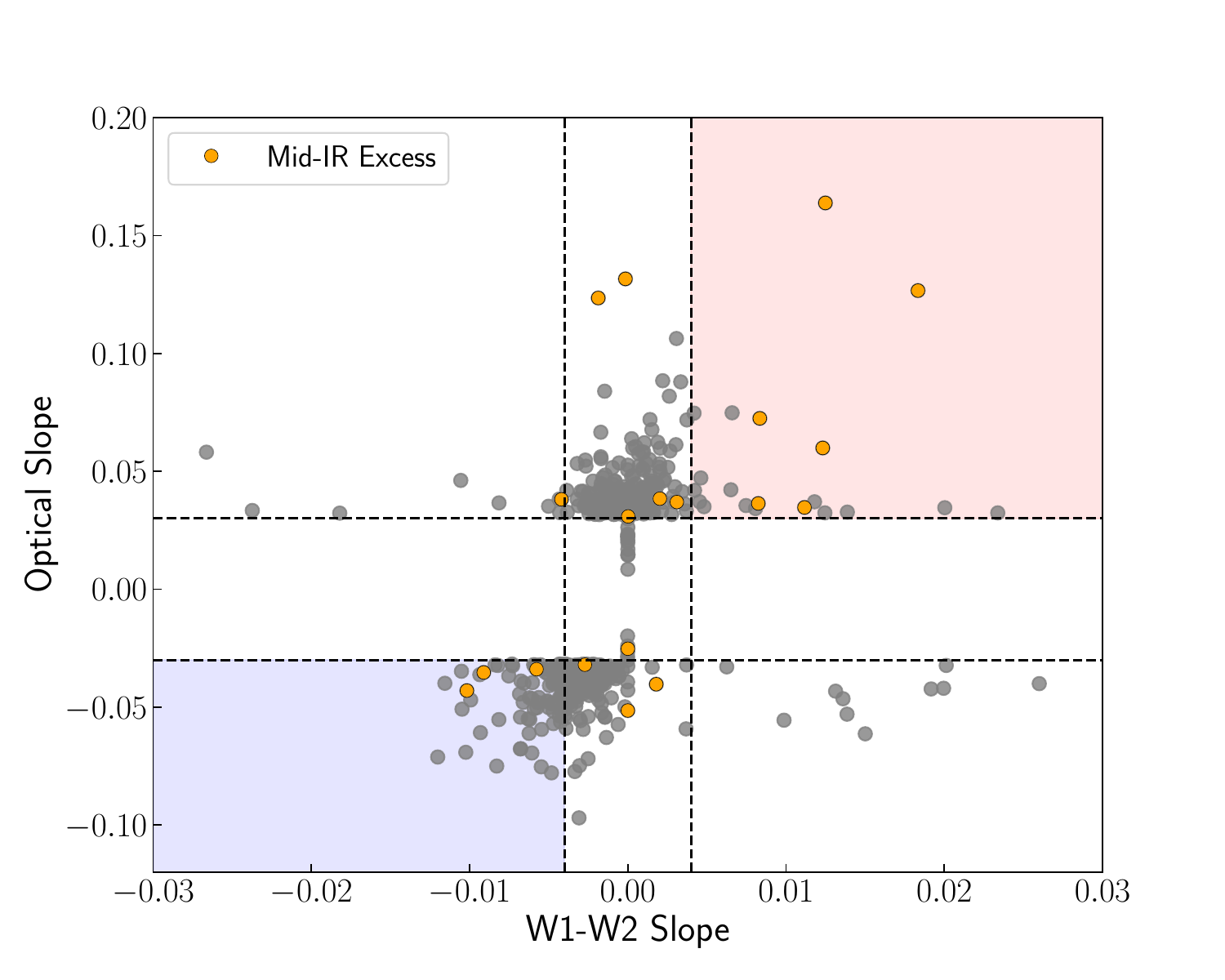}\includegraphics[width=\columnwidth]{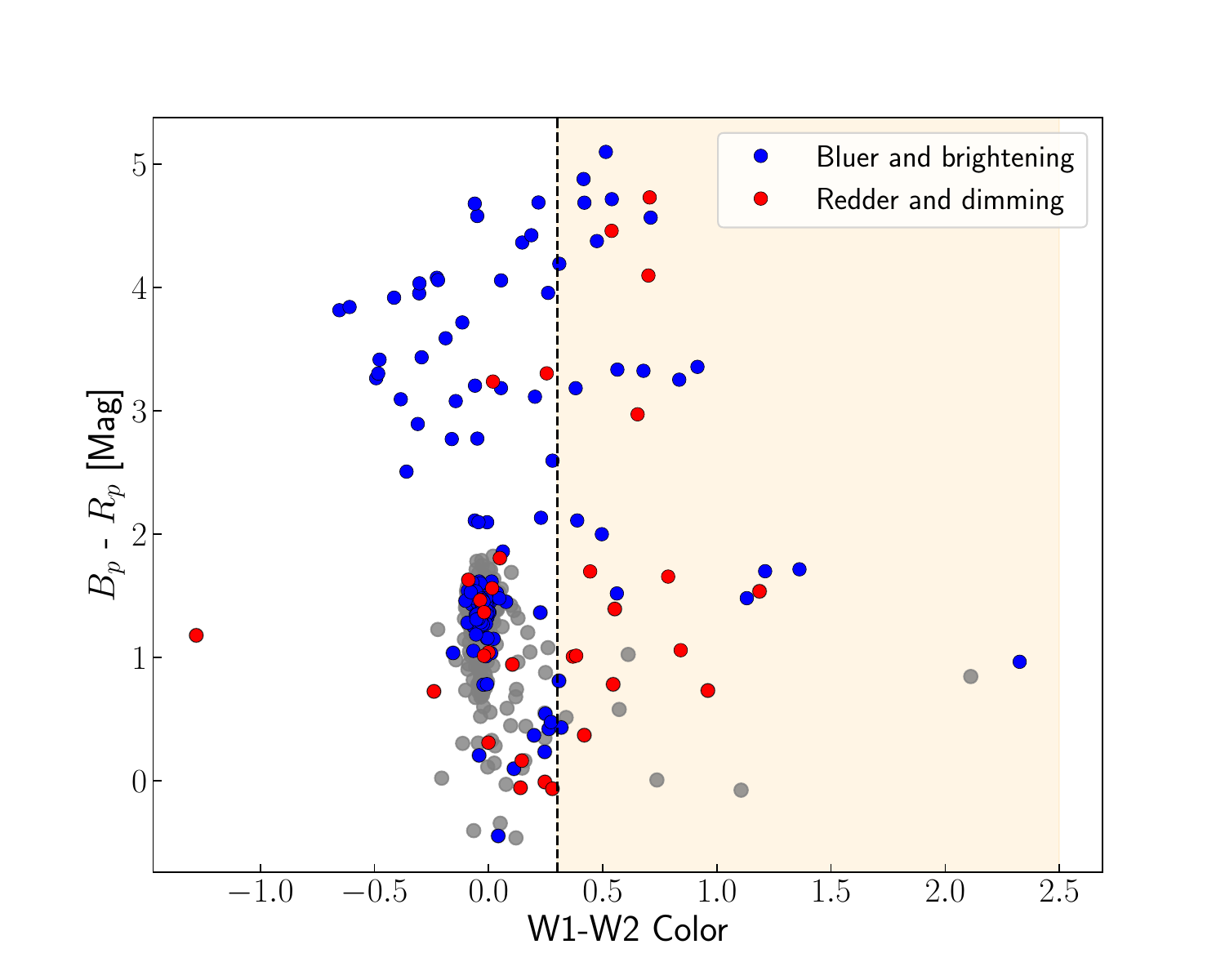}
    \caption{Left: Optical variability slope as a function of the W1$-$W2 color variability slope. Three stars are outside the edge of the figure, where 2 had bad WISE light curves and the third is the Nova V0339 Del (see Fig.~\ref{fig:cat}). Red and blue shaded regions indicates candidates with W1$-$W2 $>$ 0.3 mag that are likely getting redder and dimmer, or bluer and brighter. Right: $G_{BP} - G_{RP}$ color as a function of W1$-$W2 color. Candidates in the orange shaded region have a mid-IR excess due to dust emission. Candidates from the shaded region of each panel are shown with colored symbols on the other panel.}

    \label{fig:wise}
\end{figure*}

\begin{figure*}
    \includegraphics[width=2\columnwidth]{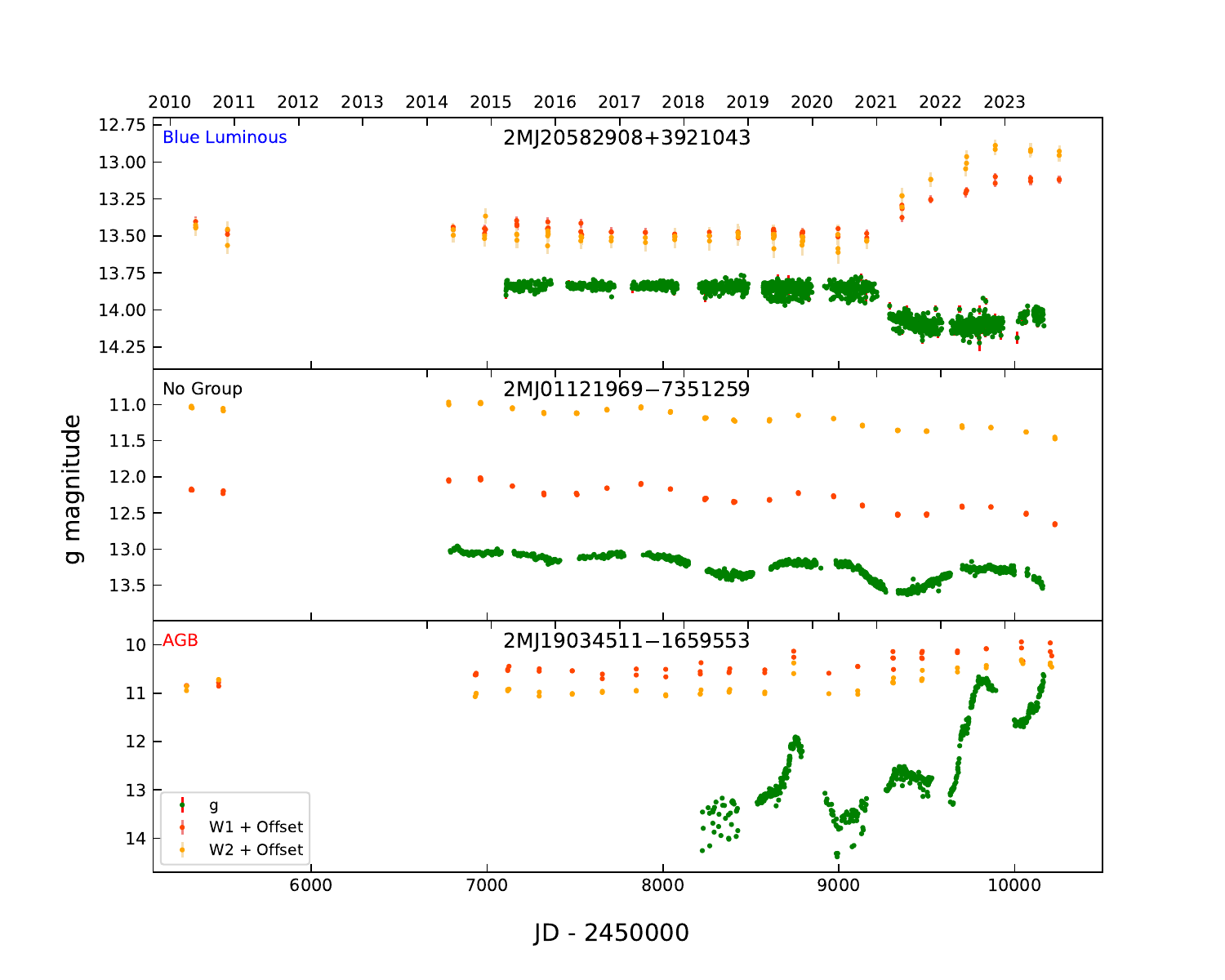}
    \caption{Comparisons of ASAS-SN and WISE light curves of candidates labeled by their 2MASS ID and CMD group. The top panel shows a candidate which clearly shows variability due to dust formation. The middle panel shows an example with very similar optical and mid-IR variability and a significant mid-IR excess. The bottom panel is the symbiotic star V919~Sgr, with a very large ($\sim 3$~mag) optical brightening and a very small mid-IR brightening.}
    \label{fig:wise2}
\end{figure*}

\begin{figure*}
    \includegraphics[width=2\columnwidth]{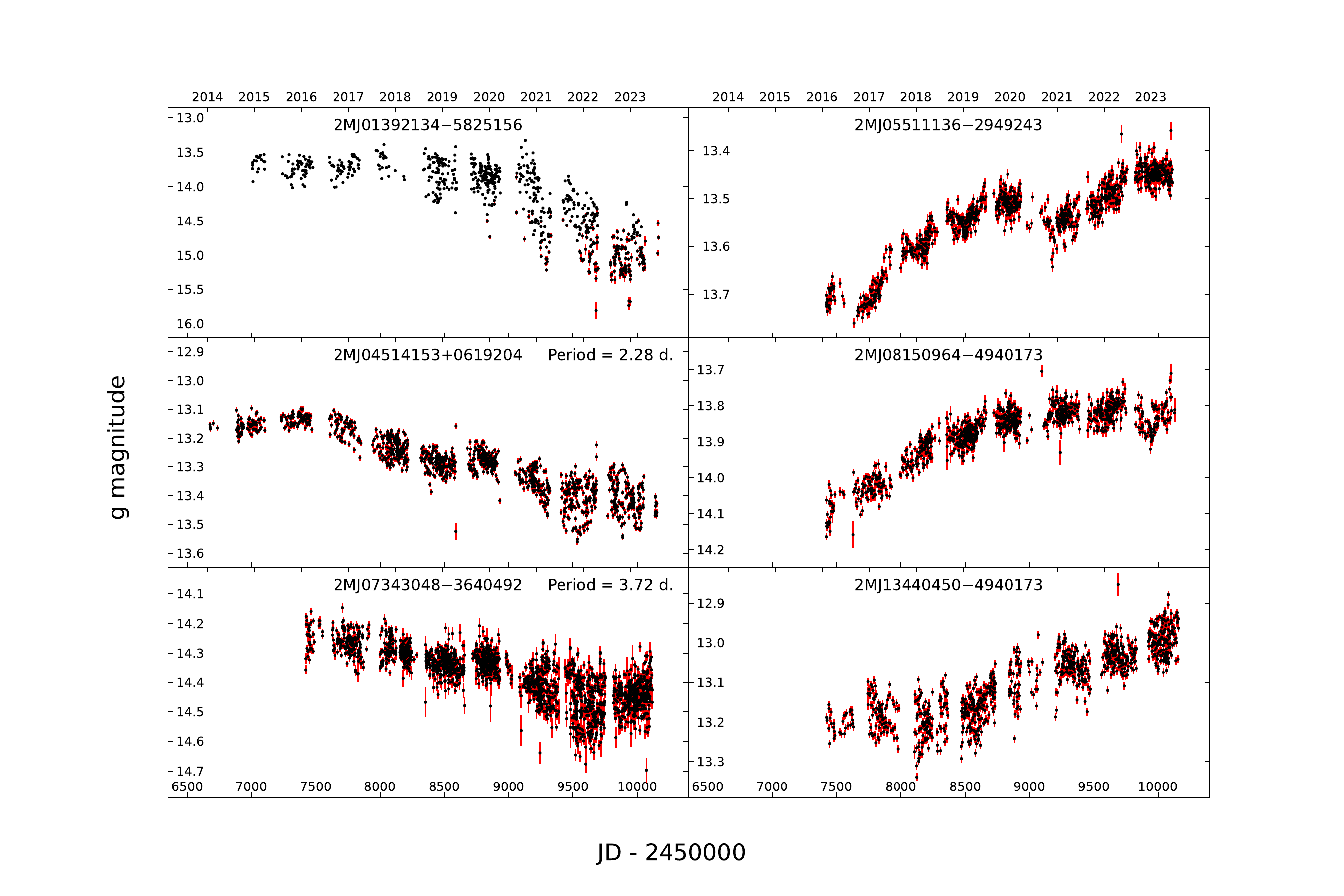}
    \includegraphics[width=2\columnwidth]{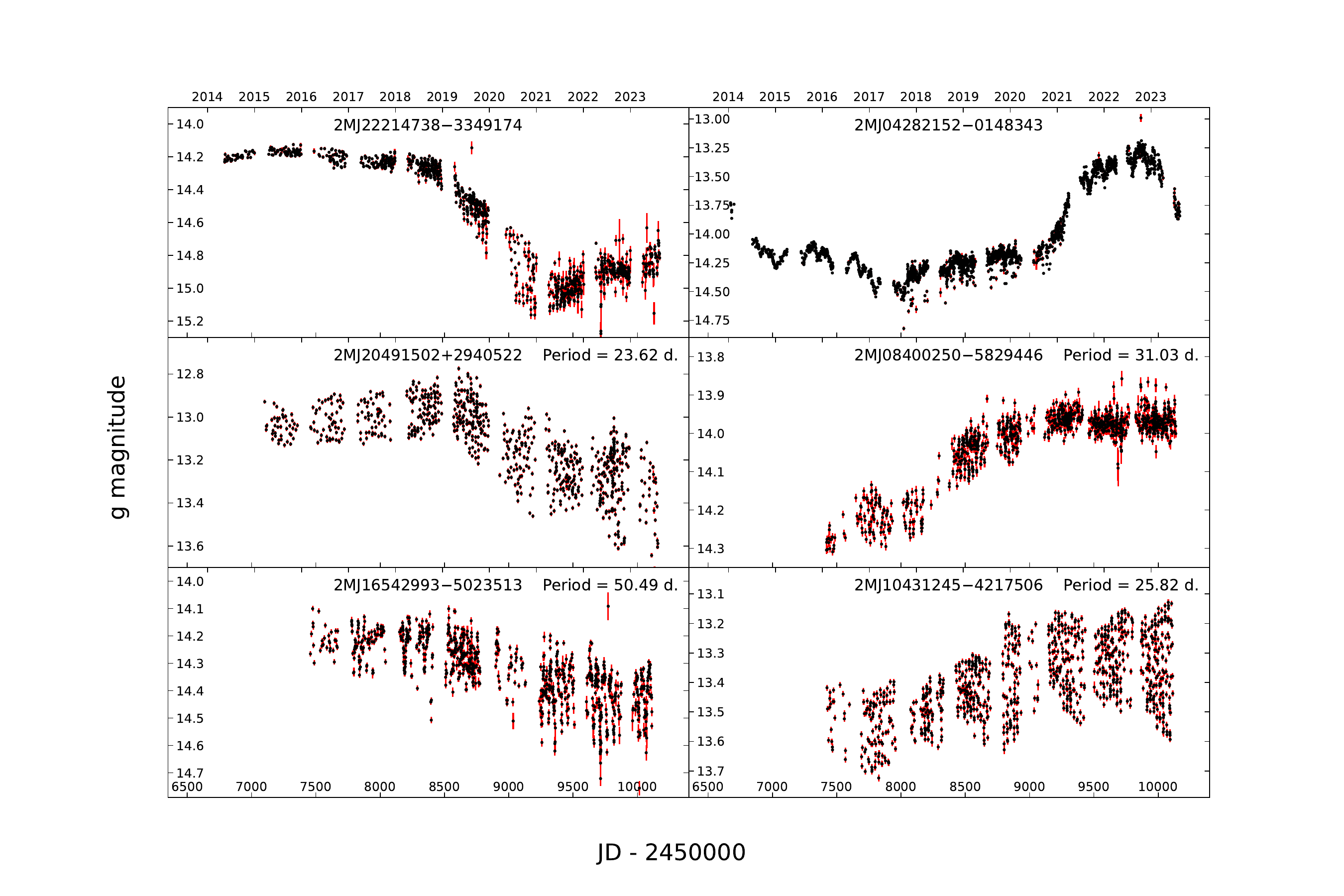}\\
    \caption{Example light curves for the main sequence group (top) and subgiant/giant group (bottom) labeled by their 2MASS ID. The left (right) panels show the 3 dimming (brightening) sources with the largest (top), median (middle), and smallest (bottom) slopes in the sample. The time axes are the same, but the magnitude ranges vary by object. If a period exists, it is given in the upper right corner.}
    \label{fig:mainseq}
\end{figure*}



\begin{figure*}
    \includegraphics[width=2\columnwidth]{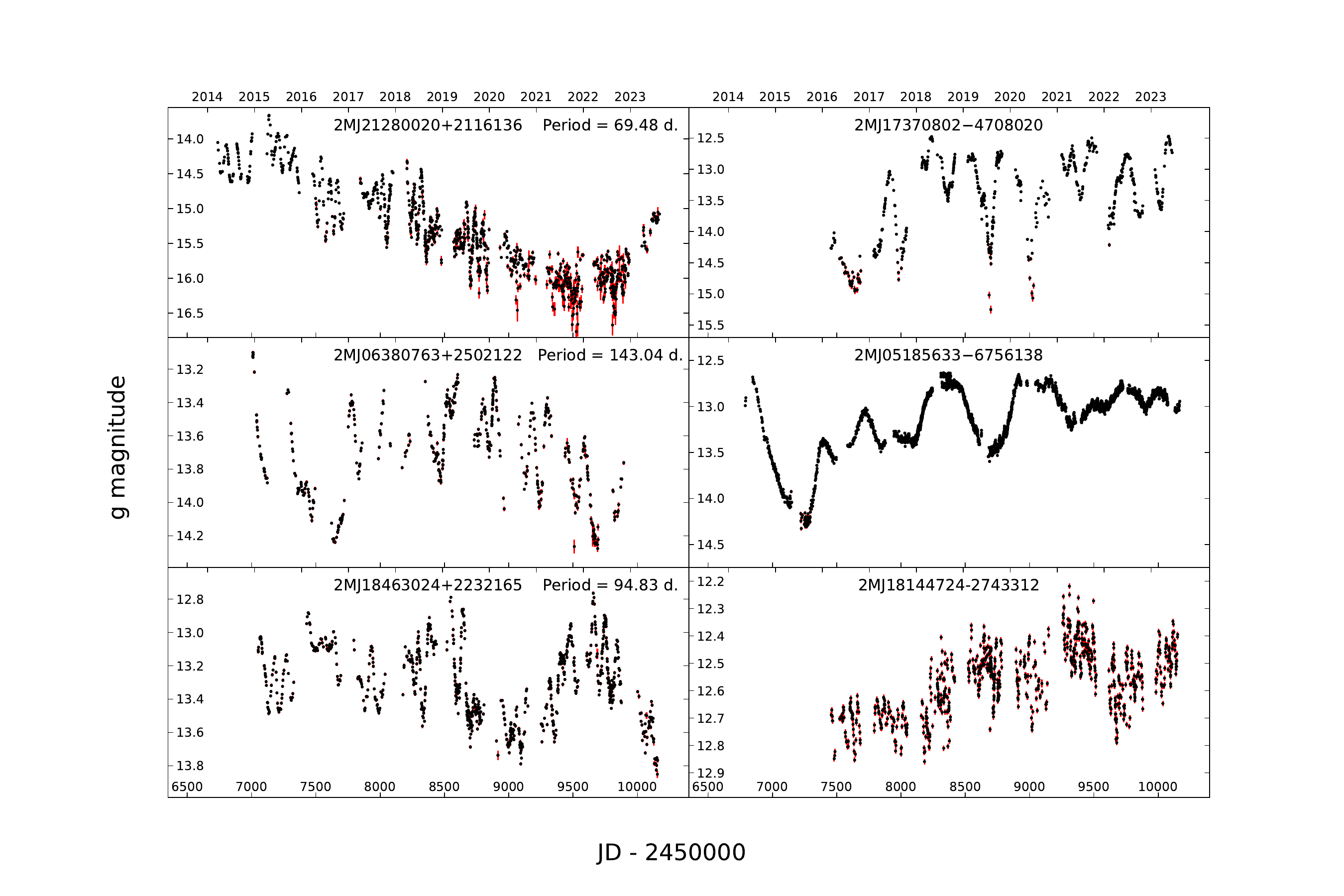}
    \includegraphics[width=2\columnwidth]{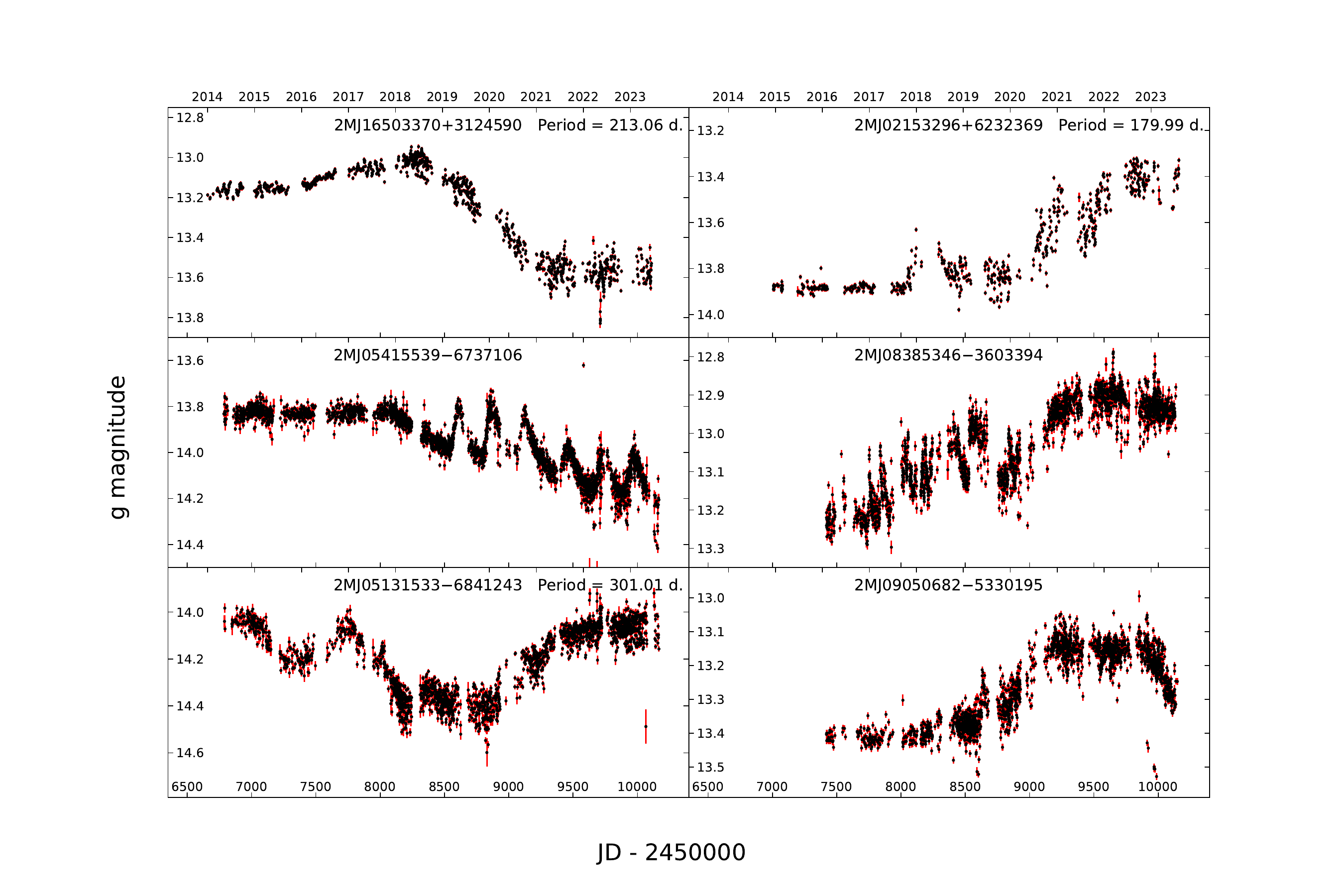} \\
    \caption{Example light curves for the AGB group (top) and luminous blue group (bottom). The format is the same is in Fig.~\ref{fig:mainseq}.}
    \label{fig:rsg}

\end{figure*}



\begin{figure*}
    \includegraphics[width=2\columnwidth]{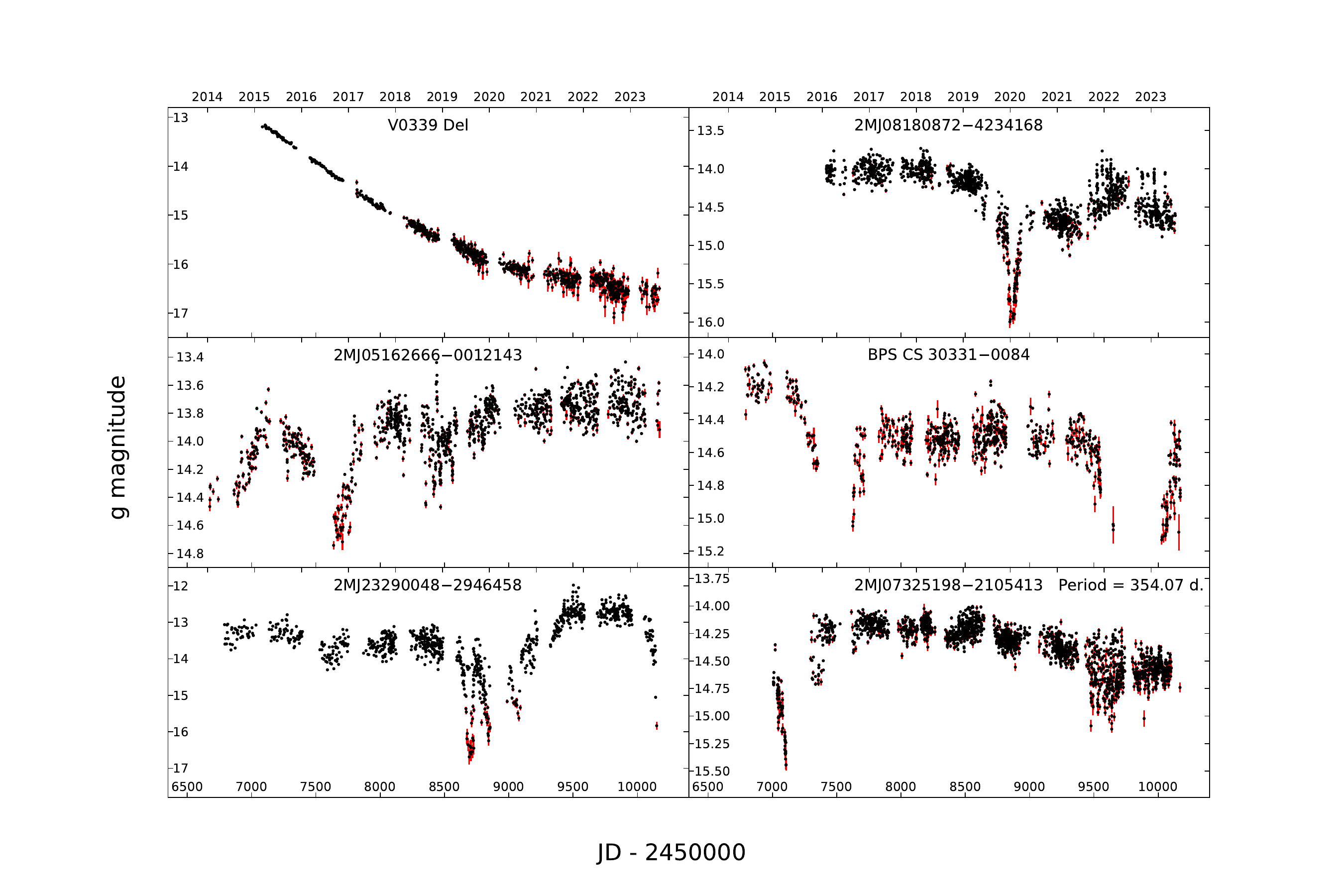}
    \includegraphics[width=2\columnwidth]{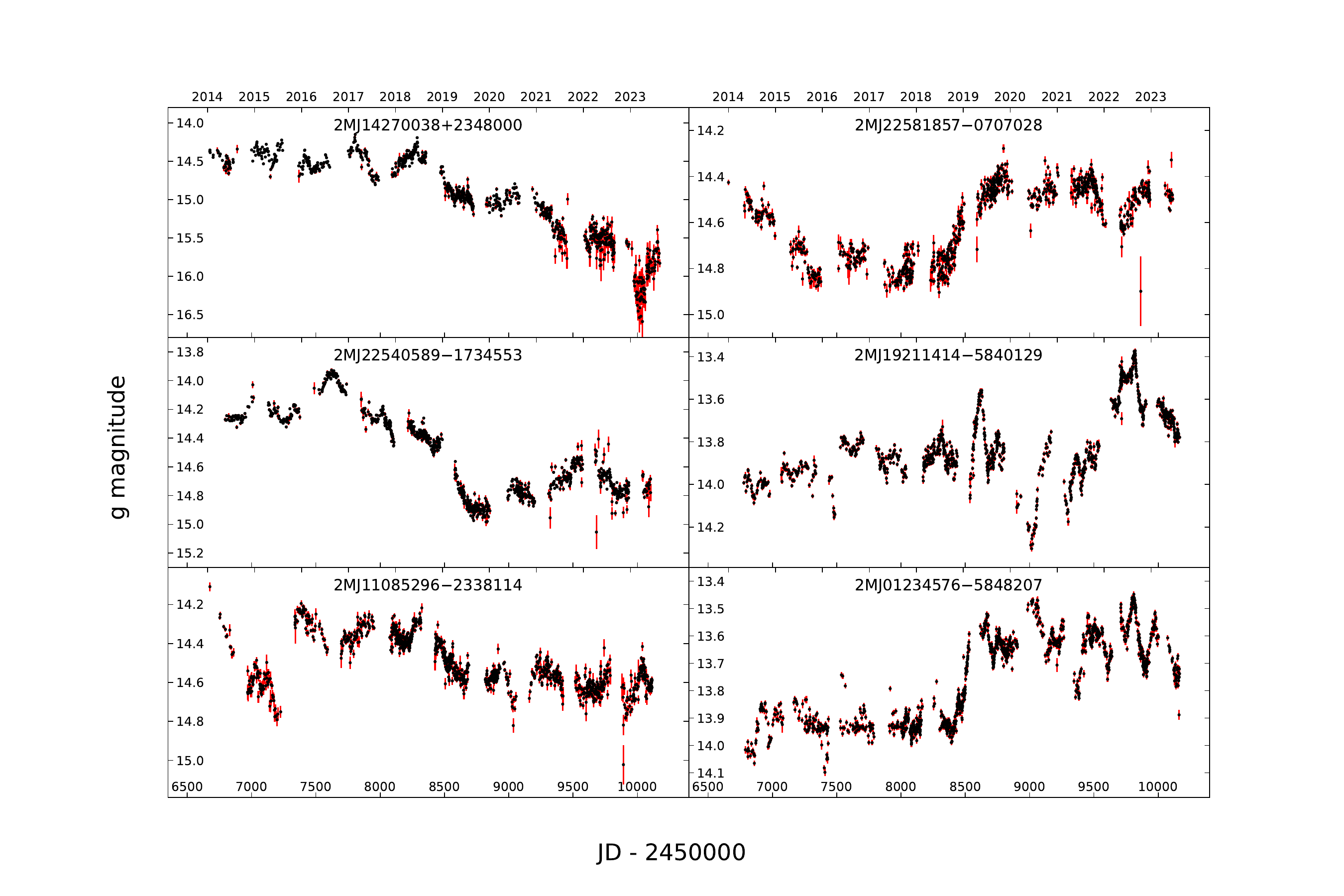} \\
    \caption{Example light curves for the novae group (top) and AGN (bottom). The source in the top left novae panel, V0339 Del, is simply a fading classical nova. The format is the same is in Fig.~\ref{fig:mainseq}.}
    \label{fig:cat}

\end{figure*}



An important question is the extent to which the optical variability is associated
with dust.  Fig.~\ref{fig:wise} shows two comparisons of the optical and WISE 
mid-IR properties of the candidates.  The right panel compares the optical
extinction corrected $B_P-R_P$ color to the mean WISE $[W1]-[W2]$ color.  We
exclude stars with $[W1]<9$ mag where saturation starts to affect the
colors.  In this space, stars redder than $[W1]-[W2] \sim 0.3$~mag are
good candidates for stars with circumstellar dust because they have a significant mid-IR excess.  
The left panel compares the optical G-band light curve slope $s_G$ and the mid-IR $[W1]-[W2]$ color $s_{col}$.  Candidates for variability driven
by new dust formation should have $s_G>0.03$ (fading) and $s_{col}>0.004$ (becoming
redder), while variability driven by dust being destroyed or becoming
more distant from the star should have the reverse.  We flag 171 fading/brightening systems and 20 systems with a mid-IR excess as possible dust driven variability candidates, as shown in both panels in Fig.~\ref{fig:wise}.
Fig.~\ref{fig:wise2} shows examples of (1) a clear dust event, (2) a Be star in the SMC (\citealt{Heydari1990}) with a mid-IR excess but no evidence that dust drives the changes in the optical brightness, and (3) V919~Sgr, which shows large optical variability and little mid-IR response. V919~Sgr is a symbiotic variable (\citealt{Ivison1993}), so the large amplitude g-band variability is likely driven by variable accretion onto the white dwarf companion of the cool giant.
 
There are 169 candidates in the main sequence group, of which 76 show additional
periodicity. Their mean
brightness changes are fairly linear, with modest bumps and wiggles.  They are
concentrated towards the lower main sequence in the CMD (Fig.~\ref{fig:CMDgroups}),
with periods generally greater than $\sim 3$~days when periodic.  They appear
to largely lie below the Kraft break \citep{Kraft1967}, which suggests that rotation,
convection and magnetic activity are driving the variability.  
Roughly, 10 percent of the sources are listed as YSOs (11) or T Tauri stars (5) in \texttt{SIMBAD} (see Table~\ref{tab:cands}).
However, compared to all the ROT variables in this magnitude range, there are many 
fewer short period systems, as seen in Fig.~\ref{fig:twinsperiods}. 
The slope amplitudes appear to simply
correspond to the high amplitude tail of the distribution for rotational variables in Fig.~\ref{fig:twinslopes}.
Their behavior does not seem to be associated with 
dust since only a few have red mid-IR colors or significantly
opposite signs for the slopes of the optical flux and the mid-IR color. Fig.~\ref{fig:mainseq}
shows six examples of light curves, where we select the three fading sources and three brightening
sources with the minimum, median, and maximum amplitudes based on the larger of the slope and $\Delta g/10$~years$^{-1}$. Additionally, we choose sources that have light curves that span at least 9 years.
The typical example has a relatively steady trend extending over the decade of observation, but almost all show evidence of an extremum, which suggests that this is also roughly the timescale for starting to revert to the mean. 
There are stars with more abrupt shifts, such as the example in the top left panel, where the star maintains a roughly constant magnitude for the first 5 years and then fades by $\sim 2$~mag over the next 4 years.

The subgiant/giant group consists of 334 stars, nearly half of the final candidates.
Stars in this group exhibit behavior similar to that of the main sequence group, 
typically having monotonically increasing and decreasing light curves.
However, they tend to exhibit larger variability amplitudes in each season, and a larger fraction are periodic (273 of 334). The period distribution
of the candidates is similar to that of the overall ROT/SR variables in this
part of the CMD.  
In the Gaia CMDs,
they lie towards the red edge of the $B_P-R_P$ color distribution both
relative to all stars (Fig.~\ref{fig:CMDgroups}) and stars classified
as rotational variables (Fig.~\ref{fig:twincmd}).  They roughly lie
in the region occupied by RS~CVn and (sub)sub-giant
rotational variables (see \citealt{Anya2024}).
Like the main sequence
candidates, their amplitudes appear to simply be the extreme tail of the
overall distribution (Fig.~\ref{fig:twinslopes}).
65 candidates are flagged for dust related variability.  
A few are classified in \texttt{SIMBAD} (see Table~\ref{tab:cands}) as YSOs (3) and R CorBor stars (1).
Fig.~\ref{fig:mainseq} shows 6 example light
curves.
The typical example for this group generally has a steady trend over the decade of observation with shorter periodic variability. Like the main sequence group, many show evidence of an extremum. The top panels display examples of stars with more extreme shifts. 
The example in the top left has a steady trend over its first 4 years, followed by a dimming event of $\sim 1$~mag spanning several years and a period of slow brightening. 
The example in the top right slowly dims and brightens over 6 years before abruptly brightening by 1 mag over a period of 2 years before again beginning to fade quickly.

The 203 candidates in the AGB group show distinct differences from the 
first two groups. They tend to have large amplitude variability within
each season, and almost all show signs of periodicity (184 of 203).
Their periodicity is generally SR-like, and
their period distribution is similar to the overall variable population (Fig.~\ref{fig:twinsperiods}).
This group contains all 25 stars flagged in {\tt SIMBAD} as Carbon stars (see Table~\ref{tab:cands}).
They are again simply the high amplitude tail of the slope distribution
(Fig.~\ref{fig:twinslopes}).  As noted earlier, they tend to be modestly
more luminous than both randomly selected stars in that part of the CMD
(Fig.~\ref{fig:CMDgroups}) and known L/SR/ROT variables in that region  (Fig.~\ref{fig:twincmd}). 
Cross-matching these stars to \texttt{SIMBAD}, we find that 5 of them are cataloged as R Coronae Borealis variables (see Table~\ref{tab:cands}).
Three candidates from this group have a large mid-IR excess and one candidate classified as an R CrB star appears to have dust-related variability.
Fig.~\ref{fig:rsg} shows 6 example light
curves, selected in the same way as for Fig.~\ref{fig:mainseq}.
The typical examples of this group have frequent, mostly irregular variability and slow brightness trends with amplitudes often larger than the first two groups and much longer periods. 
The examples with the largest overall long term variability are in the top panels. 
The top left panel has a star with some periodicity that fades $\sim 2.5$~mag over 8 years before beginning to brighten. 
The top right panel contains a star with lots of short-term variability within its observing seasons, including multiple dimming events.
In addition to these events, it increases its median brightness $\sim 2$~mag over the total duration of its light curve.

The blue luminous group is one of the smallest groups, containing, 43 candidates. 
This group mainly contains stars with sudden jumps or dips in their light curves, 
which are very different from the trends displayed in other groups. 
There are 17 periodic variables, but the remaining candidates generally are not 
periodic and exhibit the most stochastic behavior of the entire sample. 
Many of the candidates appear to have eruptive variability similar to the behavior of Be stars.  We cross matched the stars with the Be Star Spectra Database (BeSS, \citealt{BeSS}) and found only one match, but 7 are Be (1) or emission line (6) stars in \texttt{SIMBAD} (see Table~\ref{tab:cands}).
It was not obvious what existing variable class to use for comparisons of amplitudes and periods. Many of this group (18 out of 43 or $\sim 42$ percent) of this group were flagged as having dust driven variability such as the example in Fig.~\ref{fig:wise2}.
Fig.~\ref{fig:rsg} shows 6 example light
curves, selected in the same way as for Fig.~\ref{fig:mainseq}.
Typical examples for this group contain more abrupt brightening or dimming events and show clear evidence of extrema, much like the first two groups.
Their amplitudes of variability are sometimes more extreme, mirroring the amplitudes of the AGB group. 
The most extreme examples of variability in this group are shown in the top panels.
The star in the top left panel maintains a relatively constant brightness for 4 years before rapidly fading and brightening over several years and then quickly dimming by $\sim 3$~mag. This dimming event lasts for $\sim 1$~year before the star begins to rapidly brighten once again.

The novae group sits to the left of the main sequence and
contains only 6 candidates. This group is made up entirely of previously 
classified cataclysmic variables. One, nova V0339 Del \citep{Shore2013}, is a 
fading classical nova.  The rest are ``anti-dwarf'' novae (NL/VY variables
in AAVSO) showing short, deep dimming episodes.
The fadings in the anti-dwarf novae are theorized 
to be caused by low rates of mass transfer from their luminous white dwarf companions
\citep{Warner1995,Kato2002}. 
The amplitudes of this group appear to be larger than other CVs in AAVSO.
Only one star from this sample had any periodicity, and two of the novae were flagged as having dust related variability.
Fig.~\ref{fig:cat} shows all 6 as ordered in the earlier figures.
Excluding V0339 Del, the typical behavior of this group is characterized by dimming events of various lengths.

We find that 8 of our final candidates are known AGN. 
These AGN have variability within each observing season and throughout the total duration of their light curves.
Fig.~\ref{fig:cat} shows 6 example light
curves.

The most dramatic brightness changes are seen in the top left panel, where the AGN PKS~1424+240 fades by $\sim 2$~mag over 10 years.  PKS~1424+240 is a TeV-detected blazar (\citealt{Acero2015}) at a redshift of $0.60$ (\citealt{Paiano2017}).  2MJ22581857$-$0707028 (RXSJ22583$-$0707) is a moderate ($z=0.21$) redshift
radio quasar, while the remaining 6 
(2MJ0519$-$3239/ESO362$-$18, 2MJ0123$-$5848/ESO113$-$45, 2MJ1921$-$5840/ESO141$-$55, 2MJ0652$+$7425/IC450, 2MJ1108$-$2338/HE1106$-$2321, and 2MJ2254$-$1734/MR2251$-$178) all have $z<0.1$.

\begin{table*}
\centering
\caption{Long-term variability candidates and their properties. The full table is available in the electronic version of the paper. The classifications and spectral types are from \texttt{SIMBAD}.}
\scriptsize
\begin{tabular}{p{2.7cm} S[table-format=1.3] S[table-format=1.3] S[table-format=1.2] l S[table-format=1.3] S[table-format=1.2] S[table-format=1.2] l l}
\hline
2MASS & {$g$} & {Optical Slope (mag/yr)} & {$\Delta g$} & Group & {Period (days)} & {W1$-$W2 Color} & {W1$-$W2 Slope} & Classification & Spectral Type \\ 
\hline
2MJ13390896$-$0318071 & 13.00 &         0.041 &   0.373 & Subgiant/giant &  19.704 &   -0.056 &            0.001 & \ldots & \ldots \\
2MJ20324096$+$4114291 & 13.02 &         0.035 &   0.282 &            AGB &  33.905 &        0.700 &            0.012 &    \ldots       & B3-4Ia+ \\
2MJ18035783$-$2425349 & 13.03 &      -0.088 &   0.756 &         \ldots & 199.814 &      0.562 &          -0.020 &  \ldots & K0\\
2MJ08371244$-$2846184 & 13.03 &         0.066 &   0.516 &  Blue Luminous &  60.243 &   -0.409 &         -0.028 & \ldots & \ldots \\
2MJ18172030$-$4756384 & 13.04 &      -0.071 &   0.769 &  Main Sequence &   4.461 &   -0.023 &         -0.012 & EB & \ldots \\
2MJ09243724$-$2804347 & 13.04 &      -0.043 &   0.383 & Subgiant/giant &  \ldots &   -0.046 &         -0.004 & SB & \ldots \\
2MJ19491489$+$4221569 & 13.05 &      -0.038 &   0.306 &  Main Sequence &  \ldots &   -0.026 &         -0.003 & \ldots & \ldots \\
2MJ06412567$+$6450318 & 13.05 &      -0.063 &   0.571 &            AGB & 132.599 &       0.710 &         -0.048 & \ldots & M7 \\
2MJ01493939$+$6403322 & 13.06 &         0.045 &   0.406 &            AGB &  \ldots &      0.129 &            0.001 & \ldots & \ldots \\
2MJ16205657$+$2121449 & 13.06 &         0.035 &   0.317 & Subgiant/giant &  39.761 &   -0.039 &         -0.005 & \ldots & \ldots \\
\hline
\end{tabular}
\label{tab:cands}
\end{table*}

\section{Conclusions}
\label{sec:conclusions}
We select 782 slow variable candidates out of a sample 9,361,613 sources with 13<g<14.5 g mag.
They were chosen to have mean variability rates of $\gtorder 0.03$ mag/year over roughly 10 years. This sample includes 349 previously classified variables and 433 new variables. 
However, the existing classifications do not capture this type of variability.
We group these candidates into groups, finding that most are subgiants and giants.
We find 551 candidates to be periodic, most having periods longer than 10 days and belonging to the subgiant/giant group.
Using WISE light curves, we find 171 candidates with mid-IR brightness changes opposite to the optical, suggestive of changes in circumstellar dust and 20 candidates with a large mid-IR excess. The full list of variables is given in Table~\ref{tab:cands}
and their light curves can be obtained using ASAS-SN 
SkyPatrol 2.0 
(http://asas-sn.ifa.hawaii.edu/skypatrol/).

The main sequence candidates are below the \cite{Kraft1967} break, they have the smallest slopes of the entire sample, and mostly have short periods if periodic.
Only 20 are flagged as possibly having dust related variability. Their activity likely represents the extremes of star spot changes on longer time scales.
Ten percent of these candidates are flagged as YSOs or T Tauri stars in \texttt{SIMBAD}. The subgiants/giants fall on the redder edge of the color distribution of RGB stars and ROT/SR variables, have similar amplitudes to the main sequence group, and have mostly intermediate length periods. 
Stars in this group are likely associated with RS CVn/(sub)sub-giant rotational variables. However, $\sim19$ percent of these candidates are flagged as having dust-related variability.
The flagged AGB stars seem to be slightly more luminous than L, SR, or random variables. Generally they have high amplitude slow ($\geq$ 50 days) periodic variability as well. Only a few of of these candidates (one is a R CrB variable) are linked to a large mid-IR excess, implying that their variability is not driven by dust.

The blue stars are more eruptive, with large amplitudes of variability and little periodicity.
Their behavior seems similar to the behavior of Be stars, however only 7 are flagged as Be or emission line stars in BeSS and \texttt{SIMBAD}. A large fraction of this group is flagged as having dust related variability.
The novae group has deep fadings spanning several magnitudes and is entirely made up of cataclysmic variables.  
Finally, we find a small number of AGN (8), where the highest amplitude source is the z=0.60 blazar, PKS~1424+240.

Particularly for the systems with the
largest optical flux changes, it would be interesting to compare the spectra and spectral energy distributions of the stars in their high and low states. For cool
stars, large changes at V-band can be driven by modest changes in temperature without similarly large changes in luminosity, but this is more challenging for the high amplitude blue stars where the g-band no longer lies on the blue wing of the spectral energy distribution.
We plan to expand our list of candidates by continuing the search for slow variability in both brighter saturated stars (see \citealt{Winecki2024}) and dimmer than our current magnitude range. As the time spanned by ASAS-SN continues to grow, we plan to search for smaller variations and explore lower amplitudes of variability. Searches for still slower changes than $\sim 0.03$~mag/year will require significant improvements in false positive rejection and/or longer light curves.

\section*{Acknowledgements}

CSK is funded by NSF grants AST-2307385 and AST-2407206.
We acknowledge ESA Gaia, DPAC and the Photometric Science Alerts Team (http://gsaweb.ast.cam.ac.uk/alerts)
This research has made use of the SIMBAD database,
operated at CDS, Strasbourg, France 
We acknowledge with thanks the variable star observations from the AAVSO International Database contributed by observers worldwide and used in this research.
This work has made use of the BeSS database, operated at LESIA, Observatoire de Meudon, France: http://basebe.obspm.fr.

\section*{Data Availability}

Table~\ref{tab:cands} lists the stars discussed
in the paper.  Their light curves are available
from ASAS-SN Sky Patrol v2.0 (\citealt{Hart2023}).



\bibliographystyle{mnras}
\bibliography{example} 




\end{document}